\documentclass[fleqn,usenatbib]{mnras}

\usepackage{newtxtext,newtxmath}

\usepackage[T1]{fontenc}

\DeclareRobustCommand{\VAN}[3]{#2}
\let\VANthebibliography\thebibliography
\def\thebibliography{\DeclareRobustCommand{\VAN}[3]{##3}\VANthebibliography}

\usepackage{graphicx}	
\usepackage{amsmath}	
\usepackage{float}
\usepackage{natbib}
\usepackage{caption}
\usepackage{gensymb}
\usepackage{bm}
\usepackage[output-decimal-marker={,},exponent-product=\cdot]{siunitx}
\usepackage[utf8]{inputenc}
\usepackage{newtxtext,newtxmath}
\usepackage{hyperref}
\usepackage{mathtools}
\usepackage{pifont}

\newcommand{\cmark}{\ding{51}}%
\newcommand{\xmark}{\ding{55}}%

\newcommand{\mr}{\mathrm}

\title[High-redshift BHs]{The effects of super-Eddington accretion and feedback on the growth of early supermassive black holes and galaxies}

\author[F. Huško et al.]{
Filip Huško$^{1}$\thanks{E-mail: husko@strw.leidenuniv.nl},
Cedric G. Lacey$^{2}$,
William J. Roper$^{3}$,
Joop Schaye$^{1}$,
Jemima Mae Briggs$^{4}$,
Matthieu Schaller$^{1,5}$
\\
$^{1}$Leiden Observatory, Leiden University, PO Box 9513, 2300 RA Leiden, the Netherlands\\
$^{2}$Institute for Computational Cosmology, Department of Physics, University of Durham, South Road, Durham, DH1 3LE, UK\\
$^{3}$Astronomy Centre, University of Sussex, Falmer, Brighton BN1 9QH, UK\\
$^{4}$Astrophysics Research Institute, Liverpool John Moores University, 146 Brownlow Hill, Liverpool, L3 5RF, UK \\
$^{5}$Lorentz Institute for Theoretical Physics, Leiden University, PO box 9506, 2300 RA Leiden, the Netherlands
}

\date{Accepted XXX. Received YYY; in original form ZZZ}

\pubyear{2024}

\begin{document}
\label{firstpage}
\pagerange{\pageref{firstpage}--\pageref{lastpage}}
\maketitle

\begin{abstract}
We present results of cosmological zoom-in simulations of a massive protocluster down to redshift $z\approx4$ (when the halo mass is $\approx10^{13}$ M$_\odot$) using the SWIFT code and the EAGLE galaxy formation model, focusing on supermassive black hole (BH) physics. The BH was seeded with a mass of $10^4$ M$_\odot$ at redshift $z\approx17$. We compare the base model that uses an Eddington limit on the BH accretion rate and thermal isotropic feedback by the AGN, with one where super-Eddington accretion is allowed, as well as two other models with BH spin and jets. In the base model, the BH grows at the Eddington limit from $z=9$ to $z=5.5$, when it becomes massive enough to halt its own and its host galaxy's growth through feedback. We find that allowing super-Eddington accretion leads to drastic differences, with the BH going through an intense but short super-Eddington growth burst around $z\approx7.5$, during which it increases its mass by orders of magnitude, before feedback stops further growth (of both the BH and the galaxy). By $z\approx4$ the galaxy is only half as massive in the super-Eddington cases, and an order of magnitude more extended, with the half-mass radius reaching values of a few physical kpc instead of a few hundred pc. The BH masses in our simulations are consistent with the intrinsic BH mass$-$stellar mass relation inferred from high-redshift observations by JWST. This shows that galaxy formation models using the $\Lambda$CDM cosmology are capable of reproducing the observed massive BHs at high redshift. Allowing jets, either at super- or sub-Eddington rates, has little impact on the host galaxy properties, but leads to lower BH masses as a consequence of higher feedback efficiencies.
\end{abstract}






\section{Introduction}

Massive galaxies in the local Universe are found to be ‘red and dead', i.e.~largely devoid of significant amounts of gas and star formation, with old stellar populations (e.g.~\citealt{McKelvie2014}, \citealt{Davies2019}). A similar picture is found at higher redshifts in both observations (e.g.~\citealt{Santini2021}) and simulations (e.g.~\citealt{Lovell2023}). In order to explain the state of these galaxies, energy release from supermassive black holes (hereafter BHs), in the form of active galactic nuclei (AGN) feedback, is often invoked (e.g.~\citealt{DiMatteo2005}, \citealt{Bower2006}, \citealt{Croton2006}, \citealt{Booth2009}).

In the low-redshift ($z<1$) Universe, direct evidence of this feedback has been found in galaxy clusters in the form of cavities in X-ray emitting gas (e.g.~\citealt{Gull1973}, \citealt{Boehringer1993}, \citealt{Birzan2004}, \citealt{McNamara2005}, \citealt{Wise2007}), that are associated with relativistic jets and lobes of plasma emitting synchrotron radiation visible in radio frequencies (e.g.~\citealt{Blandford1979}, \citealt{Urry1995}). The power of the jets, inferred from the properties of the cavities, is large enough, in principle, to shut off any cooling from the ambient intracluster medium, thus leaving the central galaxies quenched (\citealt{Rafferty2006}, \citealt{Fabian2012}, \citealt{Hlavacek-Larrondo2012}, \citealt{Russell2013}, \citealt{Eckert2021}). AGN are also thought to affect their environments through direct heating of gas by radiation or radiation-driven winds (e.g.~\citealt{Murray1995}, \citealt{Tombesi2010}, \citealt{Feruglio2015}). Bright AGN are very numerous (e.g.~\citealt{Shen2020}), and their effects on the host galaxy properties are often apparent (e.g.~\citealt{Crenshaw2003}, \citealt{Tombesi2010}, \citealt{Feruglio2010}, \citealt{Fiore2017}).

The release of energy by BHs is facilitated by accretion through an accretion disc, and thus the study of AGN feedback is closely tied to accretion disc theory. \cite{ShakuraSunyaev1973} found a solution for a thin, radiatively-efficient disc (see its general-relativistic counterpart; \citealt{NovikovThorne1973}) that is to this day used to describe the state of bright AGN\footnote{Note also the recent simulations by \protect\cite{Hopkins2024}, who find a physically much different accretion disc model, which is dominated by magnetic fields, but with similar observable properties.}. Around $10$ per cent of the accreting mass-energy is released as radiation, through gradual viscuous heating and subsequent cooling in the accretion disc as the matter spirals inwards. Another solution found by \cite{NarayanYi1994} (see \citealt{PophamGammie1998} for the general-relativistic version) describes the geometrically thick, radiatively inefficient disc. In this solution, advection of both gas and magnetic fields is strong, the latter of which is initially amplified due to a dynamo effect, leading to the buildup of magnetic fields near the BH. Through the \cite{Blandford1977} process, energy is then extracted from the rotation of the BH, leading to the launching of relativistic jets. The thin disc solution is relevant at moderate accretion rates, while the thick disc is thought to exist at low accretion rates (\citealt{YuanNarayan2014}), the latter being the case for most BHs in the local Universe. 

At high (super-Eddington) accretion rates, the accretion disc is thought to be different again. In this case, the solution of a slim disc (e.g.~\citealt{Wang1999}) is valid: the disc is also fairly thick and advection dominated, but unlike at lower accretion rates, radiation pressure is dominant over gas pressure. \cite{Ricarte2023} recently showed, through general-relativistic radiation magneto-hydrodynamical (GRRMHD) simulations (see other examples: \citealt{Sadowski2014}, \citealt{McKinney2015}, \citealt{Pacucci2024}, that wind and especially jet launching in such a disc is fairly easy. The jets are launched as efficiently in the slim disc as in the thick disc, if the accretion rate is very high. The slim disc solution (and the super-Eddington state in general) is, however, not often invoked to explain the state of the observed AGN. This is because such a state is extremely rare in the local Universe, but the situation is likely much different at high redshifts, where galaxies are much smaller and denser (e.g.~\citealt{Bouwens2004}, \citealt{Roper2023}, \citealt{Chworowsky2024}), so BH fuelling could be much more efficient. Furthermore, even when an AGN is in such a state, it is hard to distinguish from the standard thin disc due to a low radiative efficiency.

Models for galaxy formation using large cosmological hydrodynamical simulations of galaxy formation have largely implemented either single-mode thermal AGN feedback, representing the effects of radiation on gas near the BH (e.g.~\citealt{Schaye2010}, \citealt{Teklu2015}, \citealt{Ni2022}), or additionally a kinetic mode at low accretion rates. The latter mode is isotropic in the case of IllustrisTNG (\citealt{Nelson2019}) or in the form of jets in other cases (e.g.~\citealt{Kaviraj2017}, \citealt{Dave2019}, \citealt{Dubois2021}, \citealt{Schaye2023}). The super-Eddington regime of accretion and feedback has historically received little attention, but recent simulations have begun testing it in more detail. For example, studies of individual objects (e.g.~\citealt{Lupi2016}, \citealt{Massonneau2023}, \citealt{Sassano2023}, \citealt{Bennett2024}, \citealt{Pacucci2024}) have found that allowing super-Eddington accretion may facilitate the easier growth of BHs at high redshift and allow simulations to potentially better agree with observations that are finding more massive BHs earlier than thought possible (e.g.~\citealt{Pacucci2023}). The large cosmological simulations Massive-Black (\citealt{Khandai2015}) and ASTRID (\citealt{Ni2022}) allowed accretion up to twice the Eddington rate, and \cite{Rennehan2024} have recently tested the super-Eddington regime (with no cap) using large cosmological volumes.

In the local Universe, the relation between BH mass and the host galaxy's stellar or bulge mass holds important information on BH growth and AGN feedback (e.g.~\citealt{Kormendy2013}, \citealt{McConnell2013}). With the launch of the JWST, the question of BH growth and the evolution of this relation at very high redshifts has received attention (\citealt{Ding2023}, \citealt{Maiolino2024}, \citealt{Li2024}). These studies have found that BHs may be much more massive in relation to their host galaxies (than their counterparts in galaxies of equal mass in the local Universe). However, as pointed out in detail by \cite{Pacucci2023} and \cite{Li2024}, one needs to be careful in interpreting the observed BH mass$-$stellar mass relation. The observed relation is heavily biased towards brighter objects with sufficiently wide broad lines in the AGN spectrum to obtain a reliable mass estimate. As shown by these two studies, however, the \textit{intrinsic} BH mass$-$stellar mass relation can be obtained from the observed one, given some assumptions about the scatter in the intrinsic stellar and BH populations and measurement uncertainties. \cite{Pacucci2023} find the intrinsic relation at very high redshift to be higher in normalisation than the local one, while \cite{Li2024} find one that is in good agreement with the local relation. The difference between the two studies, which use almost the same data, is that they use different assumptions about which BHs are observable; the former assuming that BHs need to have a mass of $M_\mathrm{BH}>10^{6.2}$ M$_\odot$, while the latter assume they instead need to be bright enough ($L_\mathrm{bol}>10^{44.1}$ erg~s$^{-1}$). 

In addition to BH masses, the question of BH accretion rates at very high redshift has received considerable interest, since it is hard to explain how BHs can become so massive in such a short time. Rough accretion rate estimates can be obtained from observations using the bolometric luminosity of the AGN, but such estimates assume that the radiative efficiency is of order $0.1$, while it is likely much lower for super-Eddington accretion (\citealt{Madau2014}). This in turn means the estimated accretion rates are biased against super-Eddington values. Recent simulations have successfully reproduced observations by allowing super-Eddington accretion (\citealt{Bennett2024}, \citealt{Lupi2024}, \citealt{Pacucci2024}). \cite{Lupi2024_obs} have recently pointed out that interpreting many of the BHs as being super-Eddington also brings down their BH mass estimates, thereby reducing possible tension with the local BH mass$-$stellar mass relation. There is also observational evidence of strong jet activity from AGN at high-redshifts (e.g.~\citealt{Roy2024}, \citealt{Slaus2024}), which may be relevant for galaxy and BH evolution, and which indicates that jets likely launched at super-Eddington accretion rates (\citealt{Ricarte2023}).


To simulate high-redshift galaxies observable by JWST (as well as the AGN they host), it is important to simulate a volume large enough to contain the comparatively rare and dense regions of the Universe that begin forming galaxies early. Due to the resolution and volume requirements, this leads to prohibitively large calculations. However, this expense can be circumvented by instead using zoom simulations (as we do in this work). The First Light and Reionisation Epoch simulations \citep[henceforth FLARES,][]{Lovell21, Vijayan21} used the EAGLE model \citep{Schaye2015} in conjunction with the zoom approach to simulate 40 regions that are 30 comoving Mpc (hereafter cMpc) in diameter, ranging in environment from the most underdense to the most overdense in a (3.2 cGpc)$^3$ parent volume. These 40 regions were weighted based on overdensity to ensure that the simulated galaxies represented the entire parent volume. In \cite{Wilkins24} FLARES was used to study the BH population at $z\geq5$, finding a population of massive AGN ($M_{\rm BH} > 10^9 \mathrm{M}_\odot$) at $z=5$ biased towards rare overdense environments in addition to lower mass AGN in less extreme environments. Many of these massive AGN were found to be accreting at the Eddington limit. This population is in fair agreement with pre-JWST observations but underpredicts the AGN luminosity function at $z>5$ compared to contemporary JWST studies \citep[e.g.][]{Kokorev24, Shen2020}. Additionally, in \cite{Lovell2023} a sample of passive galaxies was found in FLARES at $z=5-8$. All passive galaxies in this sample were found to contain a BH which had recently undergone a phase of extreme BH accretion and thus strong feedback, leading to the heating and evacuation of star-forming gas and the quenching of the galaxy. This highlights the importance of the co-evolution of galaxies and their AGN even at the earliest times.


Recently there has also been much interest in the Little Red Dot (LRD) population of galaxies at high redshift (e.g.~\citealt{Akins2024}, \citealt{Baggen2024}, \citealt{Kokorev2024}). LRDs are galaxies that exhibit red colors and extremely compact morphologies (with their size being of order of the point spread function of the observations). While it has been widely accepted that high redshift galaxies are much more compact than their low redshift counterparts \citep[e.g.][]{Oesch10, Kawamata15, Roper2022, Yang22, Ormerod24}, LRDs are distinct in their extremely red colors. There is considerable disagreement surrounding what drives the redness of LRDs. Some authors have proposed that the LRD population can be split into two; in one sample the colors can be explained by stellar emission and dust alone, while in the other AGN emission could explain the redness \citep[][]{Akins24, Kokorev24}. This tension highlights LRDs as an important test sample for AGN models at high redshift.

In \cite{Husko2022_spin_driven} we presented a model for BH spin evolution and AGN jet launching from the thick disc accretion regime, applicable to small accretion rates (Eddington ratios). This model was tested using a set-up of idealized galaxy groups and clusters. In \cite{Husko_winds} we expanded the model to also include BH spin evolution in the thin accretion disc (applicable to moderate, sub-Eddington accretion rates). Using the same set-up of idealized groups and clusters, we then compared these two accretion regimes and the effects of their kinetic jet (thick disc) and thermal isotropic (thin disc, representing quasar-driven winds) feedback. We found that kinetic jet feedback is more efficient at preventing or damping cooling flows in galaxy groups and clusters, leading to smaller cold gas content and star formation rates (SFRs). The entropy profiles also show that more heating is occuring at larger distances in the case kinetic jets are used. A hybrid model, where kinetic jets were used at low Eddington ratios and and thermal isotropic feedback at high Eddington ratios, seemed most promising.

Here we expand the model further by introducing the BH spin and AGN feedback in the slim disc, applicable to super-Eddington accretion rates. Some aspects of the model (the launching of jets) are also applicable to the thin, sub-Eddington disc. We apply the model to zoom-in simulations of a massive protocluster, focusing on high redshifts ($z>4$), when accretion onto the central BH is more likely to occur in the slim disc regime. We intepret our results in the context of recent observations by JWST.

In \S~\ref{sec:sec2} we present the BH spin evolution and AGN feedback model for the slim disc in detail. In \S~\ref{sec:setup} we discuss the numerical code we use to perform the simulations, as well as other numerical and physical aspects of the set-up. In \S~\ref{sec:results} we discuss the results, focusing separately on aspects related to BH growth, AGN feedback and the impact on the host galaxy and its surroundings. In \S~\ref{sec:conclusions} we summarise and conclude.

\section{Black hole evolution and AGN feedback in the slim disc}
\label{sec:sec2}

BH accretion discs have qualitatively different properties depending on the accretion rate through the disc and onto the BH. For the purpose of this paper, we expect the properties of the accretion disc around the BH to change once the accretion rate exceeds the Eddington rate
\begin{equation}
\dot{M}_\mr{Edd}=\frac{L_\mr{Edd}}{\epsilon_\mr{r}c^2}=4\pi\frac{G M_\mr{BH}m_\mr{p}}{\epsilon_\mr{r}\sigma_\mr{T}c}.
\label{eq:eq1}
\end{equation}
Here, $m_\mr{p}$ is the proton mass, $\sigma_\mr{T}$ the Thomson cross-section and $\epsilon_\mr{r}$ is the radiative efficiency of a thin, \cite{NovikovThorne1973} accretion disc, that depends on the BH spin.

If the accretion rate onto the BH exceeds the Eddington rate, we use the model that we present in the rest of this section, describing the accretion as proceeding through a slim disc\footnote{There is further super-Eddington physics that we do not include in our model. In particular, we do not include the direct effects of radiation pressure on either gas or dust. Furthermore, if radiation pressure is acting on dust, the effective Super-Eddington limit is lower and could thus affect our results (e.g.~\protect\citealt{Arakawa2022})}. For most of our assumptions, we use results of GRRMHD simulations by \cite{Ricarte2023}, and the self-similar model of \cite{Wang1999} for others. If the accretion rate is smaller than the Eddington rate, we instead describe accretion as proceeding through the thin disc. The modeling for this regime is based on the ‘standard’ accretion disc (\citealt{ShakuraSunyaev1973}), and is described in detail in \cite{Husko_winds}. We use the thin disc model even if the Eddington ratio (see the next subsection for the exact definition) $f_\mr{Edd}<0.01$, when we should use the thick disc model (\citealt{Husko2022_spin_driven}). We do this for simplicity, and we also find that using the thick disc model makes no difference here (see \S~\ref{sec:simulations}), since the BHs in our simulations are rarely at such low accretion rates, or remain in that regime for short times.

One of the main reasons we have developed these accretion models is to model BH spin, on which feedback efficiencies (and directions) are directly dependent, especially for AGN jets. The BH spin, $a$, is a dimensionless parameter that is related to the angular momentum of the BH, $J_\mr{BH}$, through its definition $a=J_\mr{BH}c/M_\mr{BH}^2G$. Its values lie in the range $a\in[-1,1]$. We limit the magnitude of spin to $0.998$ (see \citealt{Thorne1974}). The sign of spin indicates whether accretion is prograde (positive) or retrograde (negative), relative to the BH spin direction.

\subsection{Feedback efficiencies}
\label{sec:feedback_effs}

We assume that BHs surrounded by a slim disc release energy through three different channels: jet, radiation, and winds. The winds may be launched partly by radiation pressure, but we assume that at least some part of the radiation escapes the accretion disc region and thus does not drive an accretion disc wind. 

The power released through a given channel, $P_i$, is related to the accretion rate onto the BH, $\dot{M}_\mathrm{BH,acc}$, through the relation
\begin{equation}
    P_i = \epsilon_i\dot{M}_\mathrm{BH,acc}c^2,
\label{eq:feedback_eff_definition}
\end{equation}
where $\epsilon_i$ is the efficiency of energy release. In effect, the above relation is a definition for the efficiency parameter. 

The accretion rate $\dot{M}_\mathrm{BH,acc}$ in this paper is assumed to be equal to the large-scale accretion rate (in other words, we assume no mass loss to winds or other processes as the matter accretes from large scales, through the disc and onto the BH). 

\subsubsection{Jet efficiency}
\label{sec:feedback_effs_jet}

Based on simulation results by \cite{Ricarte2023}, the jet efficiency in the slim disc can be described in the same way as that for the thick accretion disc (at very low accretion rates, see e.g.~\citealt{Tchekhovskoy2010}). The jets are launched on account of the \cite{Blandford1977} (BZ) process, and their power depends on BH spin and the strength of the magnetic field. The jet power is given by $P_\mr{jet}=\epsilon_\mr{jet}\dot{M}_\mr{BH}c^2$, and we use the jet efficiency formula from \cite{Tchekhovskoy2010}:
\begin{equation}
    \epsilon_\mr{jet}=\frac{\kappa}{4\pi}\phi_\mr{BH}^2\Omega_\mr{BH}^2\big(1+1.38\Omega_\mr{BH}^2-9.2\Omega_\mr{BH}^4\big),
\label{eq:epsilon_jet}
\end{equation}
where $\kappa=0.05$ is a numerical factor that depends on the initial geometry of the magnetic field and $\Omega_\mr{BH}=a/2r_\mr{H}$ is the dimensionless angular velocity of the event horizon, with $r_\mr{H}=1+\sqrt{1-a^2}$ the dimensionless radius of the horizon. $\phi_\mathrm{BH}$ is the dimensionless magnetic flux threading the BH horizon, and it is given by
\begin{equation}
\phi_\mathrm{BH} = \frac{(f_\mathrm{Edd}/1.88)^{1.29}}{1+(f_\mathrm{Edd}/1.88)^{1.29}}\phi_\mathrm{thick,MAD}.
\label{eq:phi_slim}
\end{equation}
Here, $f_\mathrm{Edd}=\dot{M}_\mathrm{BH,acc}/\dot{M}_\mr{Edd}$ is the Eddington ratio (note that it is defined using the accretion rate onto the BH, rather than the rate at which the BH mass is changing) and $\phi_\mathrm{thick,MAD}$ is the magnetic flux threading the BH in a thick accretion disc where the magnetic field has saturated to the limit so that the system enters the so-called ‘magnetically arrested disc’ (MAD) state (\citealt{Narayan2003}, \citealt{YuanNarayan2014}). In this state, the magnetic field is maximal. The flux in this state is given by (\citealt{Narayan2021})
\begin{equation}
    \phi_\mr{thick,MAD}(a)=-20.2a^3-14.9a^2+34a+52.6.
\label{eq:phi_a}
\end{equation}

The only difference between the jet efficiency of a thick and slim disc is that the flux $\phi_\mr{BH}$ is suppressed relative to $\phi_\mr{thick,MAD}$ by the Eddington ratio-dependent factor (Eqn.~\ref{eq:phi_slim}). At sufficiently high Eddington ratios (highly super-Eddington accretion), the jet efficiency becomes equal to the one in the thick disc. By adopting Eqn.~(\ref{eq:phi_slim}) we are assuming that all accretion discs are in a MAD state, be they slim or thin (i.e.~even at $f_\mathrm{Edd}<1$). In reality this is probably true for sufficiently super-Eddington slim accretion discs, because they are advection-dominated, so they are likely to result in a similar state (in terms of magnetic fields) as the thick disc, which is consistent with being MAD in observations (see e.g.~\citealt{EHT2021}, \citealt{EHTSagA5}). However, it could be that for the thin disc (and even for a slim disc that is around or slightly above the Eddington limit) the MAD assumption is too strong, and that some observed discs are in the SANE state (\citealt{Sadowski2016}). Our results should thus be treated as an upper limit in the sense of jet strengths and jet-induced spindown, in the case where the Eddington ratio does not satisfy $f_\mathrm{Edd}\gg1$.

\begin{figure*}
\includegraphics[width=\textwidth, trim = 0 10 0 0]{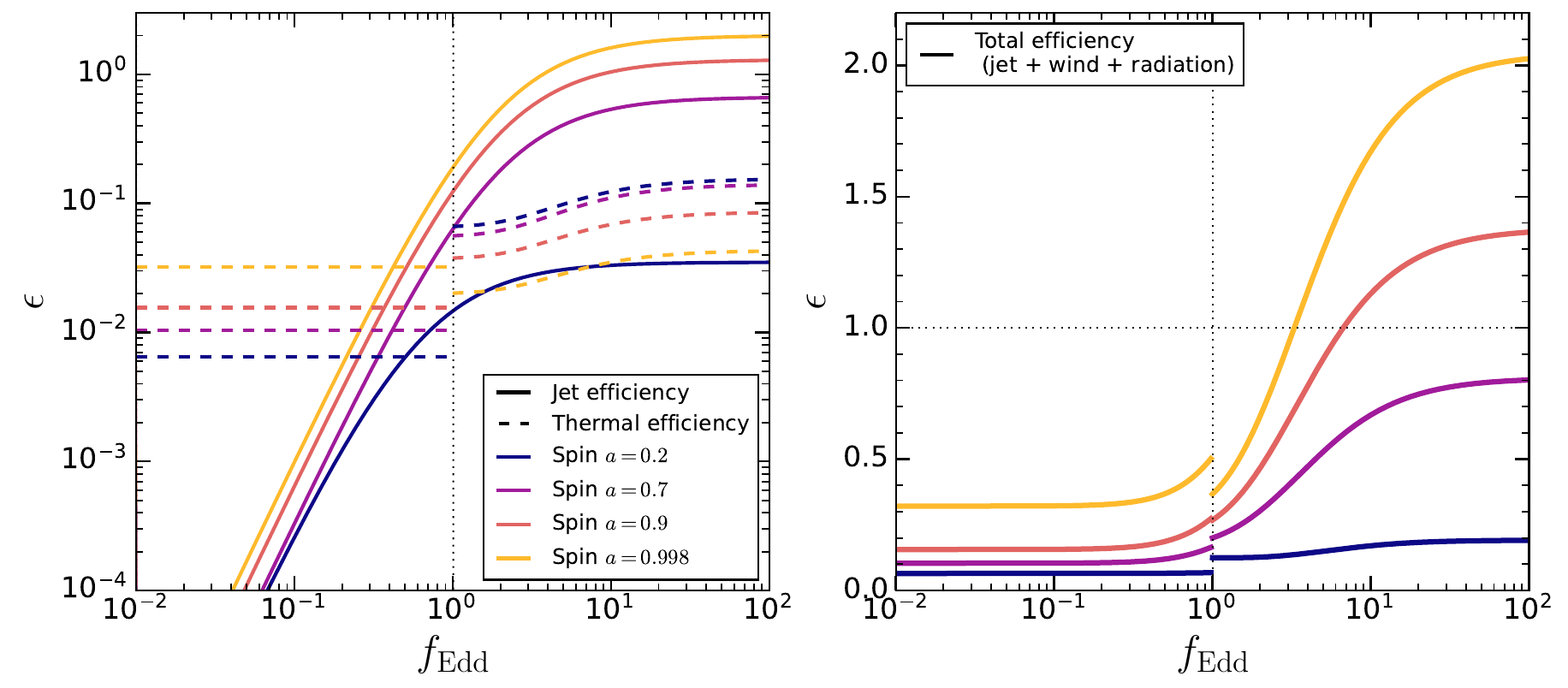}
\caption{The energy release efficiencies used in our model in both the slim accretion disc (at super-Eddington rates) and the thin disc (at sub-Eddington rates). In the left-hand panel, the solid lines show the jet efficiency given by Eqn.~(\ref{eq:epsilon_jet}), which is applicable in both regimes. The dashed lines show the efficiencies we use for thermal feedback, which we assume is driven by accretion disc winds that shock on unresolved scales. At super-Eddington rates, we assume the efficiency given by Eqn.~(\ref{eq:eps_wind_slim}), while at sub-Eddington rates, we assume that 10 per cent (the coupling efficiency) of the radiation couples on our simulated scales as thermal feedback. For that regime, we use the \protect\cite{NovikovThorne1973} radiative efficiency. In the right-hand panel, we show the sum of the wind, jet and radiative efficiencies, where the radiative efficiency at super-Eddington rates is given by Eqn.~(\ref{eq:rad_SD}). When the sum of all three efficiencies approaches or exceeds unity, the BH growth is expected to be stifled.}
\label{fig:fig0}
\end{figure*}%

In the left-hand panel of Fig.~\ref{fig:fig0}, we show the dependence of our assumed jet efficiency on the Eddington ratio for a few different values of the BH spin. The saturation at high Eddington ratios is visible, with the jet efficiency achieving values that may exceed $100$ per cent, provided that the spin is high enough. However, this saturation is relatively gradual; it occurs by $f_\mathrm{Edd}\approx10$, rather than slightly above $f_\mathrm{Edd}$. For reference, at exactly the Eddington limit ($f_\mathrm{Edd}=1$), the jet efficiency is 10 per cent of its maximum possible value (the saturated value at $f_\mathrm{Edd}\to\infty$). 

The simulations by \cite{Ricarte2023} are applicable not only to super-Eddington rates, but also to sub-Eddington ones (in the thin disc accretion regime), and we indeed use their results in the sub-Eddington regime in one of our simulations (see also simulations by \citealt{Curd2023} that find the similar possibility of efficient jet launching in sub-Eddington accretion discs). The behaviour of our assumed jet efficiency at sub-Eddington ratios is also visible in Fig.~\ref{fig:fig0}. The jet efficiency scales as $\epsilon_\mathrm{jet}\propto f_\mathrm{Edd}^{2.6}$ in this regime. At $f_\mathrm{Edd}=0.4$, for example, the jet efficiency is 2 per cent of its saturated value. For a high spin of e.g.~$a=0.998$, and the same Eddington ratio ($f_\mathrm{Edd}=0.4$), the jet efficiency is around $3$ per cent. This is comparable to the total feedback efficiency used for thermal feedback (typically of order 1 per cent, e.g.~\citealt{Schaye2015}) in this Eddington ratio regime, indicating that jets may be important even in the thin disc. In the thin disc regime, our thermal feedback efficiency is $\epsilon_\mathrm{f}\epsilon_\mathrm{rad,NT}$ (see \citealt{Husko_winds}), where $\epsilon_\mathrm{f}=0.1$ is the coupling efficiency for thermal feedback from the thin disc, and $\epsilon_\mathrm{rad,NT}$ is the spin-dependent \cite{NovikovThorne1973} radiative efficiency of a thin accretion disc. In Fig.~\ref{fig:fig0}, at sub-Eddington accretion rates, we indicate the values of this total thermal feedback efficiency, which is to be compared to the jet efficiency. The jet efficiency is larger than the thermal efficiency above $f_\mathrm{Edd}\approx0.4$ (with only a slight dependence on spin). For a smaller Eddington ratio of e.g.~$f_\mathrm{Edd}=0.1$, the jet efficiency is only $\approx0.1$ per cent, which is negligible compared to the thermal feedback efficiency in the thin disc. Thus, jets are important at $f_\mathrm{Edd}\geq0.4$, but likely unimportant at accretion rates below that in the thin disc regime (unless the Eddington ratio becomes lower than $\approx0.01$, at which point the disc begins to transition to the thick disc state and the jets dominate).

\subsubsection{Wind efficiency}

\cite{Ricarte2023} find that their accretion discs also launch winds, on account of a combined effect of radiation pressure, thermal pressure and MHD effects. They provide results for their total MHD efficiency (which includes both the jets and the winds). We estimate their wind efficiency by subtracting from their total MHD efficiency the jet efficiency given by Eqn.~(\ref{eq:epsilon_jet}), as explained in more detail in Appendix \ref{app}. We then fit with the following formula (albeit with large uncertainty) to the resultant wind efficiency values:
\begin{equation}
\epsilon_\mr{wind}(a) = 0.065\bigg[1+\bigg(\frac{\phi_\mathrm{BH}}{50}\bigg)^2\bigg] \max(0,1+\Omega_\mathrm{BH}-8\Omega_\mathrm{BH}^2\big).
\label{eq:eps_wind_slim}
\end{equation}

Note that \cite{Ricarte2023} measure their total MHD efficiency at $5R_\mathrm{G}$. While this should encompass most of the jet energy output, it may underestimate the total energy of the wind (that reaches large radii) since significant wind acceleration may be occuring at radii larger than $5R_\mathrm{G}$. The wind efficiency given by Eqn.~(\ref{eq:eps_wind_slim}) may thus be an underestimate of the total wind efficiency. We use it regardless, since it is the only formulation we are aware of that parameterizes wind efficiencies from the slim disc as a function of spin and Eddington ratio, using results from realistic GRRMHD simulations.

The assumed wind efficiency is shown in the left-hand panel of Fig.~\ref{fig:fig0} at super-Eddington rates. It is of order $2-20$ per cent and varies only by a factor of a few as the Eddington ratio increases from $f_\mathrm{Edd}=1$ to $f_\mathrm{Edd}\to\infty$. We find that the wind efficiency is larger than the jet efficiency only for $a<0.4$.

\subsubsection{Radiative efficiency}

We do not use the radiative efficiency of the slim disc to inject any energy on galaxy scales in the simulation. This is because the effects of the radiation on wind launching are already implicitly included in the wind efficiency (Eqn.~\ref{eq:eps_wind_slim}). However, we do assume a radiative efficiency for the purpose of calculating AGN luminosities in the slim disc state. We take results based on the numerical work by \cite{Sadowski2009} (instead of \citealt{Ricarte2023}, because they do not discuss their radiative efficiencies in detail). We use the following fitting formula adopted for their results by \cite{Madau2014}:
\begin{equation}
\epsilon_\mr{rad}=\frac{0.1}{f_\mr{Edd}}A(a)\bigg[ \frac{0.985}{1.6/f_\mr{Edd}+B(a)}+\frac{0.015}{1.6/f_\mr{Edd}+C(a)}\bigg],
\label{eq:rad_SD}
\end{equation}
where the three spin-dependant functions are given by $A(a)=(0.9663-0.9292a)^{-0.5639}$, $B(a)=(4.627-4.445a)^{-0.5524}$ and $C(a)=(827.3-718.1a)^{-0.7060}$.

\subsection{BH growth}

Given the efficiencies we have introduced in the preceding subsection, we may now specify the rate at which the BH grows (or reduces) its mass:
\begin{equation}
    \dot{M}_\mathrm{BH}= (1-\epsilon_\mathrm{jet}-\epsilon_\mathrm{wind}-\epsilon_\mathrm{rad})\dot{M}_\mathrm{BH,acc},
\label{eq:growth}
\end{equation}
where $\dot{M}_\mathrm{BH,acc}$ is the accretion rate through the accretion disc and onto the BH, and $\epsilon_i$ is the efficiency of a given channel $i$. The factor multiplying the accretion rate is sometimes referred to as the accretion efficiency; this is not to be confused with the accretion efficiency factor that encapsulates mass loss to winds (which we do not use in this paper). As apparent from the previous subsections, the sum of all energy release efficiencies may be quite large. In fact, it may exceed unity. In the right-hand panel of Fig.~\ref{fig:fig0}, we show the sum of all three energy release efficiencies that we use in the thin disc (at $f_\mathrm{Edd}<1$) and slim disc ($f_\mathrm{Edd}>1$) regimes. At low values of BH spin (e.g.~$a=0.2$), the total efficiency is generally negligible. 

However, if the spin becomes high enough ($a>0.7$), and the Eddington ratio high enough (factor of a few above Eddington), the total energy release efficiency approaches or even exceeds unity. In this scenario, the BH releases large amounts of energy as it accretes, but its total mass increases very little, or even decreases, due to the $(1-\epsilon)$ factor in Eqn.~(\ref{eq:growth}). In other words, the BH may lose more mass due to the tapping of the rotational component of its total mass-energy (mainly to the jets) than it gains through accretion. Its spin also rapidly reduces due to jet-induced spindown in this scenario (see the next subsection). 

\subsection{Evolving the magnitude of the black hole spin}
\label{sec:spin_magn}

To describe BH spinup/spindown, we again use results from \cite{Ricarte2023}. Their simulations include all relevant effects that impact BH spin evolution, including accretion and outflows (jets and winds) that tap some of the angular momentum of the BH. We use their fitting formulae that smoothly interpolate between the thin disc regime without significant jet feedback (for $f_\mathrm{Edd}\ll1$), and highly super-Eddington accretion in the slim disc ($f_\mr{Edd}\gg1$) where jet feedback essentially matches the thick disc (and so jet spindown should also be similar). They formulate the spinup/spindown function as
\begin{equation}
    \frac{\mr{d}a}{\mr{d}M_\mr{BH,acc}/M_\mr{BH}}=s_\mathrm{HD} + s_\mr{EM},
\label{eq:da_dlnMBH_Ricarte}
\end{equation}
where $\mr{d}M_\mr{BH,acc}=\dot{M}_\mr{BH,acc}\mr{d}t$ is an increment of accreting mass (before any energy losses due to winds, jets or radiation), and $\dot{M}_\mr{BH,acc}$ is the large-scale accretion rate (see the next subsection). In the above equation, the first term on the right-hand side is a purely hydrodynamical one, while the second is an electromagnetic term. $s_\mathrm{HD}$ is given by
\begin{equation}
    s_\mathrm{HD}=\frac{s_\mathrm{thin}+s_\mathrm{min}\xi}{1+\xi},
\label{eq:s_HD}
\end{equation}
where $\xi=0.017f_\mathrm{Edd}$, $s_\mathrm{min}=0.86-1.94a$ and $s_\mathrm{thin}=\ell_\mathrm{ISCO}-2a e_\mathrm{ISCO}$ is the spinup/spindown function of the ‘pure’ thin disc (with no outflows and outside the MAD regime), in which $\ell_\mathrm{ISCO}$ and $e_\mathrm{ISCO}$ are the (dimensionless) specific angular momentum and binding energy, respectively, at the innermost stable circular orbit (ISCO; see appendix A of \citealt{Husko2022_spin_driven}). The EM term is given by\footnote{This term is missing a minus sign in \protect\cite{Ricarte2023}.}
\begin{equation}
    s_\mathrm{EM}=-\mathrm{sgn}(a)\epsilon_\mathrm{EM}\bigg(\frac{1}{k\Omega_\mathrm{H}}-2a\bigg),
\label{eq:s_EM}
\end{equation}
where $\epsilon_\mathrm{EM}$ is the total (jet+wind) EM efficiency, and $k$ is given by 
\begin{equation}
        k=
\begin{dcases}
    0.23 ,& a < 0 \\
    \min(0.35,0.1+0.5a),              & a>0
\end{dcases}
\label{eq:k}
\end{equation}

\subsection{Deciding the sign and direction of the black hole spin}
\label{sec:spin_sign}

In order to choose the sign and evolve the direction of the BH spin, we use the same procedure as in \cite{Husko2022_spin_driven} (and we thus give only a summary here). This is motivated by the fact that both the thick disc and the slim disc are advection-dominated and have a similar geometric shape ($H/R\approx0.3$).

In general-relativistic dynamics, additional torques (\citealt{LenseThirring}, hereafter LT) cause precession of a hypothetical ring of matter at a given radius from the BH. In the thin disc regime, an accretion disc either aligns or counter-aligns (\citealt{BardeenPetterson}) out to some warp radius $R_\mathrm{warp}$ due to these LT torques. In the advection-dominated discs (the thick and slim disc), the disc instead precesses about the spin vector (e.g.~\citealt{Fragile2007}). This precession occurs on much smaller time-scales than the time-scales we are simulating, so we instead assume that the disc aligns or counter-aligns (on average) with the spin vector.

The warp radius (the radius within which communication of LT torques is efficient) is given by (see \citealt{Husko2022_spin_driven})
\begin{equation}
R_\mr{warp,adv}=R_\mr{G}\bigg(\frac{384\vert a\vert}{25(H/R)^2}\bigg)^{2/5}.
\label{eq:r_warp_adaf}
\end{equation}
For the slim disc, we take $H/R=0.3$ based on GRMHD simulations (as mentioned, this is the same value as in the thick disc). The warp radius is then no more than $R_\mr{warp,adv}\approx6R_\mr{G}$.

Our model assumes that accretion proceeds in finite increments, such that one warp mass $M_\mr{warp}$ is swallowed at a time (with $M_\mr{warp}$ the disc mass within $R_\mr{warp}$). Before $M_\mr{warp}$ is swallowed, the torques between the BH and the disc bring the system to a steady state. During this process, the magnitude of the BH angular momentum remains constant, while its direction aligns with that of the total angular momentum $\bf{J_\mr{tot}}=\bf{J_\mr{BH}}+\bf{J_\mr{warp}}$. The angle between $\bf{J_\mr{BH}}$ and $\bf{J_\mr{warp}}$ decreases with time, also resulting in the decrease in the magnitude of $\bf{J_\mr{warp}}$. 

We then assume that accreting matter is either aligned or counteraligned with respect to the new spin axis (prograde or retrograde accretion, respectively). Accretion is retrograde if
\begin{equation}
\cos \theta<-\frac{J_{\mathrm{warp}}}{2 J_{\mathrm{BH}}},
\label{eq:counteralignment}
\end{equation}
where $\cos \theta=\bf{\hat{J}_\mr{BH}}\cdot\bf{\hat{J}_\mr{d}}$ is the initial misalignment between the BH and the (large-scale) angular momentum of the disc, whose direction is $\bf{\hat{J}_\mr{d}}$ (see \citealt{King2005} for a derivation). In the case that equation (\ref{eq:counteralignment}) is not satisfied, accretion is assumed to be prograde. 

The warp angular momentum in equation (\ref{eq:counteralignment}) is calculated by integrating the product of the surface density of the slim disc and $L(R)$, the specific angular momentum at a distance $R$ from the BH, out to $R_\mr{warp}$. A similar integral (without the $L(R)$ factor) is used to calculate the warp mass $M_\mr{warp}$. We use the surface density from the self-similar slim disc solution presented in \cite{Wang1999}:
\begin{equation}
\Sigma_\mr{adv}=\frac{\dot{M}_\mathrm{BH,acc}}{2\pi R\vert v_\mr{r} \vert},
\label{eq:Sigma_adv}
\end{equation}
where $v_\mr{r}=-\alpha v_0 v_\mr{K}$ is the radial velocity. Here, $v_\mr{K}=\sqrt{M_\mr{BH}G/R}$ is the Keplerian velocity, and $v_0=1/\sqrt{5}$ is a numerical coefficient. The specific angular momentum is given by $L(R)=\Omega_0\sqrt{M_\mr{BH}GR}$, where $\Omega_0=1/\sqrt{5}$ is another numerical coefficient. We assume the accretion viscosity parameter $\alpha$ to be $\alpha=0.2$.

\subsubsection{Black hole spin evolution due to mergers}

The final spin of a BH-BH merger remnant is dependant on the mass ratio of the BHs, q = $M_\mr{BH,2}/M_\mr{BH,1}<1$, as well as the three vectors relevant for this situation: the two spin vectors, $\mathbf{a_1}$ and $\mathbf{a_2}$, of the BHs, as well as the orbital angular momentum of the system about the centre of mass, $\mathbf{L}$. We use a fitting formula specifically tailored for use in cosmological simulations (\citealt{Barausse2009}):
\begin{equation}
    \mathbf{a}_\mr{fin} = \frac{1}{(1+q)^2}(\mathbf{a}_1+\mathbf{a}_2q^2+\mathbf{l}q),
\label{eq:a_fin}
\end{equation}
where $\mathbf{l}$ is a vector whose direction is the same as that of the orbital angular momentum $\mathbf{L}$ (in the centre-of-mass frame), while its magnitude is given by
\begin{align}
|\mathbf{l}|&=\frac{s_{4}}{\left(1+q^{2}\right)^{2}}\left(\left|\mathbf{a}_{1}\right|^{2}+|\mathbf{a}|_{1}^{2} q^{4}+2\left|\mathbf{a}_{1} \| \mathbf{a}_{2}\right| q^{2} \cos \phi\right)+ \\
&+\left(\frac{s_{5} \mu+t_{0}+2}{1+q^{2}}\right)\left(\left|\mathbf{a}_{1}\right| \cos \theta+\left|\mathbf{a}_{2}\right| q^{2} \cos \xi\right)+ 
2 \sqrt{3}+t_{2} \mu+t_{3} \mu^{2}.
\end{align}
Here, $\mu=q/(1+q)^2$ is the symmetric mass ratio, and $s_4 = -0.1229$, $s_5 = 0.4537$, $t_0 = -2.8904$, $t_2 = -3.5171$, $t_3 = 2.5763$. The three cosines depend on the angles between the different vectors which play a role in the merger: $\cos \phi=\hat{\mathbf{a}_{1}} \cdot \hat{\mathbf{a}_{\mathbf{2}}}$, $\cos \theta=\hat{\mathbf{a}_{1}} \cdot \hat{\mathbf{l}}$, $\cos \xi=\hat{\mathbf{a}_{2}} \cdot \hat{\mathbf{l}}$.

Given the information available within the model, we could in principle calculate the recoil velocity of the remnant, as well as the total mass fraction lost to gravitational waves. We do not implement the former at this stage since we cannot reliably track the movement of BHs in their host galaxies due to dynamical fraction. However, we do implement the latter. We use results from similar numerical relativity simulations (\citealt{Barausse2012}) and write the final mass of the remnant as:
\begin{align}
    M_\mr{BH,fin} = &(M_\mr{BH,1}+M_\mr{BH,2})\Big\{1 - [1 - e_\mr{ISCO}(\Tilde{a})]\mu  - \\
    & 4\mu^2[4p_0+16p_1\Tilde{a}(\Tilde{a}+1)+e_\mr{ISCO}(\Tilde{a})-1]\Big\},
\label{eq:E_rad}
\end{align}

where $p_0=0.04827$, $p_1=0.01707$ and $e_\mr{ISCO}(\Tilde{a})$ is the dimensionless specific binding energy at the innermost stable circular orbit calculated using an effective spin variable defined as 
\begin{equation}
    \Tilde{a} = \frac{|\mathbf{a_1}|\cos\theta+|\mathbf{a_2}|\cos\xi}{(1+q)^2}.
\label{eq:a_tilde}
\end{equation}

\section{Methods and simulations}
\label{sec:setup}

\subsection{Numerical code and subgrid galaxy formation model}

We use the SWIFT\footnote{\href{https://swiftsim.com}{https://swiftsim.com}}, open-access simulation code (\citealt{Schaller2024}) that includes hydrodynamics, gravity, cosmology, as well as various subgrid physical processes. We use a version of the code with zoom simulation optimisations (Roper et al. in prep) which are described in brief below. The code includes the SWIFT-EAGLE galaxy formation model, an updated version of the EAGLE model (\citealt{Schaye2015}). Here we do not describe the entire model in detail, but instead give a summary. 

We assume the standard, $\Lambda$CDM cosmology (using the \cite{Planck2015} cosmological parameters). The initial conditions (see \S~\ref{sec:setup}) containing both dark matter (DM) and gas are then evolved using gravity and hydrodynamics. For the former, we refer the reader to \cite{Schaller2024}. For the latter, the SWIFT-EAGLE model uses the SPHENIX smoothed-particle-hydrodynamics (SPH) scheme (\citealt{Borrow2022}). The gas cools radiatively (using cooling tables from \citealt{Ploeckinger2020}) and begins to form stars, the latter implemented using a pressure law (where the star formation rate in a self-gravitating disc is related to the surface density of gas). Stars are assumed to form at a gas entropy floor that is imposed to avoid runaway cooling down to low temperatures ($T\ll10^4$) and high densities where physical processes are not well resolved. Stellar feedback and gas enrichment are included, representing the effects of supernovae type Ia and II. The effects of both are implemented as thermal isotropic heating of gas with a single temperature ($10^7$ K).

\subsection{Black hole physics}
\label{sec:BH_physics}

BHs are seeded in the simulations in dark matter haloes with masses $M>10^{10}$ M$_\odot$ once they been identified using a friends-of-friends (FOF, e.g.~\citealt{Davis1985}) algorithm implemented in SWIFT. We assume a seed mass of $M_\mr{BH,0}=10^4$ M$_\odot$ and an initial BH spin of $\vert a_0 \vert = 0.01$ (pointed in a random direction). The latter value is non-zero since some analytical quantities needed for the BH spin evolution model diverge at $\vert a \vert = 0$. After seeding, the BHs are ‘manually’ repositioned at every time-step to the minimum of the gravitational potential of their host system, since dynamical friction is not resolved well in the simulations (see \citealt{Bahe2022} for details of the repositioning scheme).

Once seeded, we assume that the accretion rate onto the BH is given by the Bondi-Hoyle-Lyttleton rate (\citealt{Hoyle}, \citealt{Bondi}, hereafter the Bondi rate):
\begin{equation}
\dot{M}_\mr{BH,acc}=4\pi\frac{G^2M_\mr{BH}^2\rho}{(c_\mr{s}^2+v^2)^{3/2}},
\label{eq:bondi}
\end{equation}
where $\rho$, $c_\mr{s}$ and $v$ are the density, isothermal sound speed and velocity (relative to the BH) of the gas, respectively, all of which are calculated in a kernel-weighted fashion using the nearest $\approx60$ gas particles surrounding the BH in the simulation. The BH mass growth (or loss) rate is given by $\dot{M}_\mr{BH}=(1-\epsilon_\mr{tot})\dot{M}_\mr{BH,acc}$, where $\epsilon_\mr{tot}$ is the sum of all feedback (including radiative) efficiencies. Note that in reality, the accretion disc would provide some delay and further change the accretion rate onto the BH, but this effect cannot be properly accounted for unless we know both the mass and the angular momentum of the accretion disc, which is only feasible with very-high-resolution simulations (e.g.~\citealt{Koudmani2024}). In the fiducial SWIFT-EAGLE model, the accretion rate is capped at the Eddington rate.

AGN feedback is implemented through two feedback channels: 1) thermal isotropic heating representing the effects of accretion-disc winds that shock on subgrid scales and 2) kinetic jets that represent the effects of relativistic jets. The feedback power in a given channel $i$ is calculated as $P_i=\epsilon_i\dot{M}_\mr{BH,acc} c^2$, where $\epsilon_i$ is the feedback efficiency for that channel. This energy is dumped at every time-step to two separate reservoirs. For thermal isotropic feedback, once the accumulated energy is large enough, a single particle is heated by $\Delta T = 10^9$ K (\citealt{Booth2009}). The chosen particle is the one closest to the BH. 

In the kinetic jet case, the energy is accumulated until two particles can be kicked such that their kinetic energy (measured in the frame of the BH) per unit mass is increased by $v_\mathrm{j}^2/2$, with $v_\mathrm{j}=10^4$ km s$^{-1}$ (as described in detail in \citealt{Husko_winds}). Note that this is different from increasing the momentum by $v_\mathrm{j}$ in some chosen direction, since the gas already has an initial velocity in the frame of the BH. The two particles chosen for the kicking are the two closest to the BH in the two hemispheres defined by the BH spin vector. To kick a pair of particles, a random kick direction is generated from a cone with a half-opening angle of $\theta_\mr{j}=7.5\degree$ around the BH spin axis. One particle is kicked in that direction, and the other (from the other hemisphere) in the opposite direction.

At sub-Eddington rates, in the fiducial model (without slim disc modeling and BH spin), the only form of feedback is thermal isotropic, and the feedback efficiency is given by $\epsilon = \epsilon_\mathrm{f}\epsilon_\mathrm{r}$. Here, $\epsilon_\mathrm{r}=0.1$ is an ‘average’ radiative efficiency of the thin disc, while $\epsilon_\mr{f}=0.1$ is the fraction of that radiated energy that is assumed to couple to the gas and launch a wind. In the case where we use BH spin, we modify this radiative efficiency by using the spin-dependent formula from \cite{NovikovThorne1973}, which gives efficiencies in the range $4-40$ per cent. For the slim disc, we use the wind efficiency given by Eqn.~(\ref{eq:eps_wind_slim}) for the thermal isotropic feedback channel. In the cases where we use jets, we use the jet efficiency given by Eqn.~(\ref{eq:epsilon_jet}), applicable to both sub- and super-Eddington rates. The efficiencies described above are shown in Fig.~\ref{fig:fig0}.

\subsection{Physical set-up}
\label{sec:phys_setup}


We identify objects from a parent simulation of $L=400$ cMpc with periodic boundary conditions and a mass distribution realised with $N=3008^{3}$ composite particles representing baryonic matter and DM. We choose this parent volume to guarantee the inclusion of multiple massive protocluster systems while limiting the computational cost of the parent simulation. The parent simulation's initial conditions (ICs) were created at $z=127$ and evolved to $z=0$ assuming collisionless dynamics using SWIFT. The short- and long-range gravitational forces were computed using a 4th-order fast multipole method and a particle-mesh method solved in Fourier space. The simulation's ICs were generated following the same procedure as adopted by \citet[][see their Section 2.4]{Schaye2023} for the FLAMINGO simulations, using the third-order Lagrangian perturbation software \textsc{monofonicIC} \citep{Hahn_2020} coupled to the \textsc{panphasia} \citep{Jenkins_2013} Gaussian random noise field.


We select the 6 most massive objects from the parent simulation at $z=4.28$. These objects are characterised by halo masses of $M_{200}= 2.48\times10^{13} \rm M_{\odot}, 1.62\times10^{13}M_{\odot},  1.31\times10^{13}M_{\odot},  1.21\times10^{13}M_{\odot}, 1.20\times10^{13}M_{\odot}$ and $1.18\times10^{13}M_{\odot}$. The high resolution region is defined as a sphere of radius $r=8$ cMpc from the coordinates of the minimum potential of the halo computed by VELOCIraptor (\citealt{Elahi2019}).
The high resolution region is populated with four DM particles for every gas particle, configured in a face-replicating grid similar to that adopted by \citet{Richings_2021} (note that this is different from the original EAGLE model where the same number of gas and DM particles were used). This yields gas and DM particles of approximately equal mass and minimises the spurious transfer of kinetic energy from dark matter to star particles that artificially heats galaxy discs \citep{Ludlow_2019}. The zoom initial conditions are generated with a high resolution gas particle mass of $m_{\rm gas}=1.84\times10^{6} \rm M_{\odot}$ and dark matter particle mass of $m_{\rm DM}=2.42\times10^{6} \rm M_{\odot}$. The remainder of the parent volume is populated with pressureless composite particles representing baryons and DM, whose mass increases with distance from the high-resolution region.

We generate multi-resolution `zoomed’ initial conditions embedded within the parent volume using the second-order Lagrangian perturbation theory software \textsc{ic\_2lpt\_gen} \citep{Jenkins_2010}, coupled to \textsc{panphasia}. The \textsc{ic\_2lpt\_gen} software assumes the cosmic matter density is comprised only of baryons and DM, therefore the zoomed ICs are generated with cosmological parameters that differ very slightly from those of the parent box, subsuming the matter density of massive neutrinos into that of the dark matter. The cosmological parameters of the zoomed ICs are therefore: $\Omega_{\rm CDM}=0.256, \Omega_{\rm b}=0.0486, \Omega_{\Lambda}=0.695, h=0.681, \sigma_8=0.807$.

We resimulated six individual haloes using the zoomed-in ICs and the zoom-optimized code. Our aim was to find a halo whose central galaxy and BH exhibit relatively regular evolution between the different models we are interested in testing (see \S~\ref{sec:simulations}). All six haloes experience mergers as they evolve, as do their central galaxies and BHs. However, in two of the most massive haloes, some of the mergers involved more than two galaxies, with some of the simulations predicting a flyby of one of the members, and others a complete merger. This rendered the different simulations (using the same initial conditions) of limited use for our purposes. In two other simulations, the most massive galaxy (and BH) was not the primary of the most massive halo throughout the simulation, but only merged with it later on. This again rendered these initial conditions of limited use, since the evolution was harder to interpret. In the fifth halo, significant galaxy and BH growth only started very late, at $z\approx5.5$, which was very close to the redshift down to which we are running our simulations ($z=4.28$). 

\begin{center}
\begin{figure}
\includegraphics[width=0.88\columnwidth, trim = 0 10 0 0]{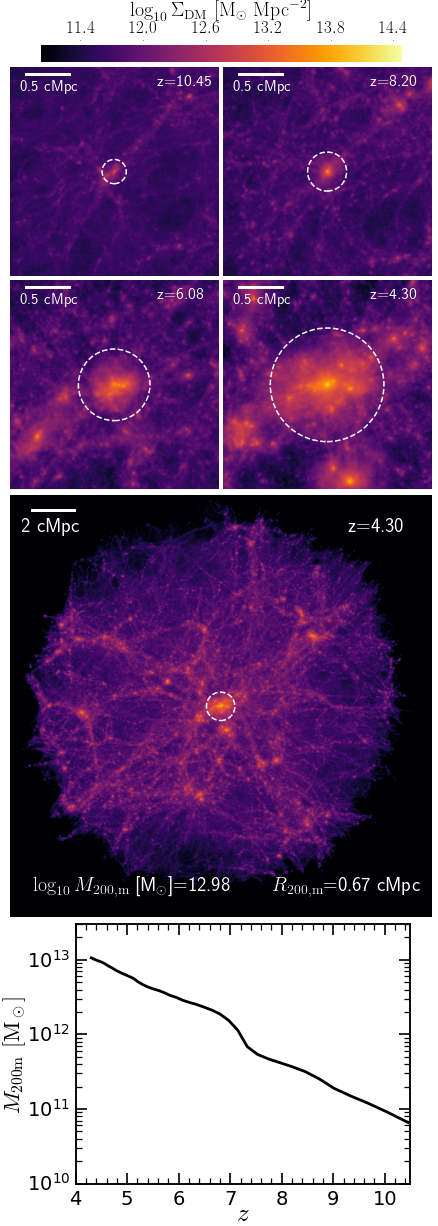}
\caption{The formation of the dark matter halo (the protocluster) hosting the BH and galaxy of interest. The first four panels show the surface density of the dark matter and are centred on the most massive, primary halo at different redshifts. The largest, middle panel shows an image of the entire zoom-in simulation region at the final time. The dashed circles indicate the extent of the virial radius (also labeled, alongside the halo mass, in the large panel). The bottom panel shows the redshift evolution of the halo mass. The data shown here corresponds to the ‘Base model’ simulation, but these plots are almost identical for the other three simulations (see \S~\ref{sec:simulations}.)}
\label{fig:fig1}
\end{figure}%
\end{center}

Given the above considerations, we opted to study in detail the sixth most massive halo in the parent simulation volume. This object was not affected by any of the issues mentioned above. In Fig.~\ref{fig:fig1} we illustrate the evolution of this primary halo in the zoom-in region. We show it both explicitly (bottom panel) and in terms of the DM surface density (other panels). This halo goes through a merger at $z\approx7.5$, which is also reflected heavily in the galaxy evolution, as well as its BH growth, since it leads to the first super-Eddington episode in this halo. A further much weaker merger event occurs at $z\approx5.5$ (note that it is not visible in the DM halo mass evolution shown in Fig.~\ref{fig:fig1}), again triggering some super-Eddington accretion onto the most massive BH.


\subsection{Zoom-in code}
The version of SWIFT used for the simulations presented in this work includes optimisations for zoom simulations. These optimisations are described in detail in Roper et al. (in prep), here we provide a summary (ignoring any modifications to the partitioning scheme which we do not use in our simulations).

SWIFT employs task-based parallelism. To facilitate this it splits the domain of the simulations into a uniform grid of "top-level" cells; each of which are given tasks (individual steps in the calculation, e.g. particle-particle gravity, kick, drift, etc.) which can be distributed over the available threads. This framework yields excellent scaling and efficient use of computational resources, however, in a zoom simulation this framework leads to a huge imbalance of work between the top-level cells which contain the zoom region and those that do not. 

To overcome this inherent imbalance, Roper et al. (in prep) introduces a secondary "top-level zoom" grid of cells isolated to the zoom region, alongside optimisations to the short- and long-range gravity interactions including this secondary grid. This seemingly simple change enables a far better division of the high-resolution zoom region and thus a better balancing of the work between tasks (/threads). 

When running a zoom simulation, hydrodynamics and sub-grid related tasks are isolated to the zoom region. This makes them efficient and simplifies the implementation, but also yields a pressureless boundary at the edges of the zoom region. To avoid long-running neighbour searches at this boundary we enforce a physically motivated maximum smoothing length (0.5 cMpc for our simulations). Any particles that leave the zoom region and enter the background are converted to dark matter. To ensure this pressureless boundary has no effect on our simulations, we included a sufficiently large padding region around the halo of interest such that its effect is entirely absent in our results.

\subsection{Structure finding}

Since we are only interested in the primary BH and its host galaxy in the simulations, we do not run a halo finder. We instead find the most massive BH at $z=4.3$, and track its most massive progenitor back to earlier times. We find that in all four of the simulations we have done, which are described in detail below, this BH always tracks the same object and remains at the centre of the primary object, which is itself found and defined as follows. At every snapshot, we find all particles within a 100 kpc physical aperture around the BH, and run an FOF with a linking length of 0.2 times the mean particle separation among all those particles (roughly corresponding to an overdensity of $200$ times the mean density). We then remove all particles that belong to subgroups and not to the main group (the latter of which is the most massive FOF group). This procedure does not perfectly eliminate all satellite structures, but we find that it works sufficiently well for our purposes. The choice of centring our apertures on the BH is further motivated by the primary interest of this study being the effects of the BH on its environment (both immediate and large-scale). We find that this procedure also always locates the central galaxy of the most massive halo in the simulation, which is itself at the centre of the zoom-in region (note that this is by design; see the previous subsection for discussion of the choice of initial conditions). Having completed the above procedure, we compute all galaxy-related properties (e.g.~total stellar mass, stellar half-mass radius, cold gas mass, etc.) in 30 pkpc physical apertures around the BH.

\subsection{Simulations}
\label{sec:simulations}

\captionsetup[table]{skip=0pt} 
\begin{table*}
\begin{center}
\caption{List of simulations performed. For each simulation, we turn on different parts of our full model, as described in the table. Note that the simulation ‘+Super-Eddington jets’ features not only the addition of super-Eddington jets, but also thermal isotropic feedback due to super-Eddington winds, as well as use of BH spin, which also affects the sub-Eddington regime. However, the latter two effects are subdominant to the addition of jets.}
\label{tab:tab1}
\end{center}

\begin{tabular*}{1.\textwidth}{@{\extracolsep{\fill}}|c|cccc|}
  \hline
   Simulation &  Super-Eddington accretion & BH spin & Super-Eddington jets and winds & Sub-Eddington jets  \\
  \hline 
  Base model & \xmark & \xmark & \xmark & \xmark \\
  +Super-Eddington & \cmark & \xmark & \xmark & \xmark \\
  +Super-Eddington jets  & \cmark & \cmark & \cmark & \xmark \\
  +Thin disc jets  & \cmark & \cmark & \cmark & \cmark \\
  \hline
\end{tabular*}

\end{table*}

We perform four simulations: ‘Base model', ‘+Super-Eddington', ‘+Super-Eddington jets’ and ‘+Thin disc jets'. These simulations are summarized in Table \ref{tab:tab1}. In the first of these, we do not use BH spin, AGN jets or any aspects of the slim disc model (\S~\ref{sec:sec2}). This corresponds to the fiducial set-up of the SWIFT-EAGLE model. For the second simulation, we allow super-Eddington accretion (without any cap), but do not change anything else. For the third simulation, we turn on modeling of BH spin and AGN jets. For this simulation, BH spin is modelled using the thin accretion disc model as in \cite{Husko_winds} at sub-Eddington rates, and the slim disc model presented in \S~\ref{sec:sec2} for super-Eddington rates. In the former regime, feedback is thermal isotropic (using a spin-dependent radiative efficiency, unlike in the base model). In the latter regime, there is both thermal isotropic feedback (due to slim disc winds) and kinetic jet feedback, but the latter is dominant for almost all values of BH spin. For the final simulation, we extend the fitting functions for the jet efficiency (Eqn.~\ref{eq:epsilon_jet}) and spinup/spindown (Eqn.~\ref{eq:da_dlnMBH_Ricarte}) from the super-Eddington to the sub-Eddington regime (instead of the jet efficiency being equal to 0 in the thin disc). This allows realistic jet launching in the thin disc (but much weaker than in the slim disc), as well as incorporating the effects of those jets on spinup/spindown in the thin disc. 

We have also performed a simulation where the thick disc is used to model BH spin and launch AGN jets at low Eddington ratios (the model presented in \citealt{Husko2022_spin_driven}). We found no discernible differences compared to using the thin disc model (and no jets) at such Eddington ratios, so we do not include the results of this simulation in the present work. The lack of differences is partly related to the fact that the BH accretion rate in these simulations rarely dips below $f_\mathrm{Edd}=0.01$, and partly to the simulation duration not being very long in terms of cosmic time, so any effects in this accretion regime may not have time to accumulate.

\section{Results}
\label{sec:results}

We begin by analyzing the growth of the primary BH in the four simulations we have run (\S~\ref{res:bh_growth}). We then discuss the feedback the BHs are doing in the different simulations in both quantitative and qualitative terms (\S~\ref{res:bh_feedback} and \S~\ref{res:bh_feedback_large_scale}). In \S~\ref{res:galaxies}, we discuss the impact of this feedback on galaxy growth and properties. Finally, in \S~\ref{res:JWST} we compare our simulations with JWST observations.

\subsection{Black hole growth}
\label{res:bh_growth}

\begin{figure*}
\includegraphics[width=\textwidth, trim = 0 10 0 0]{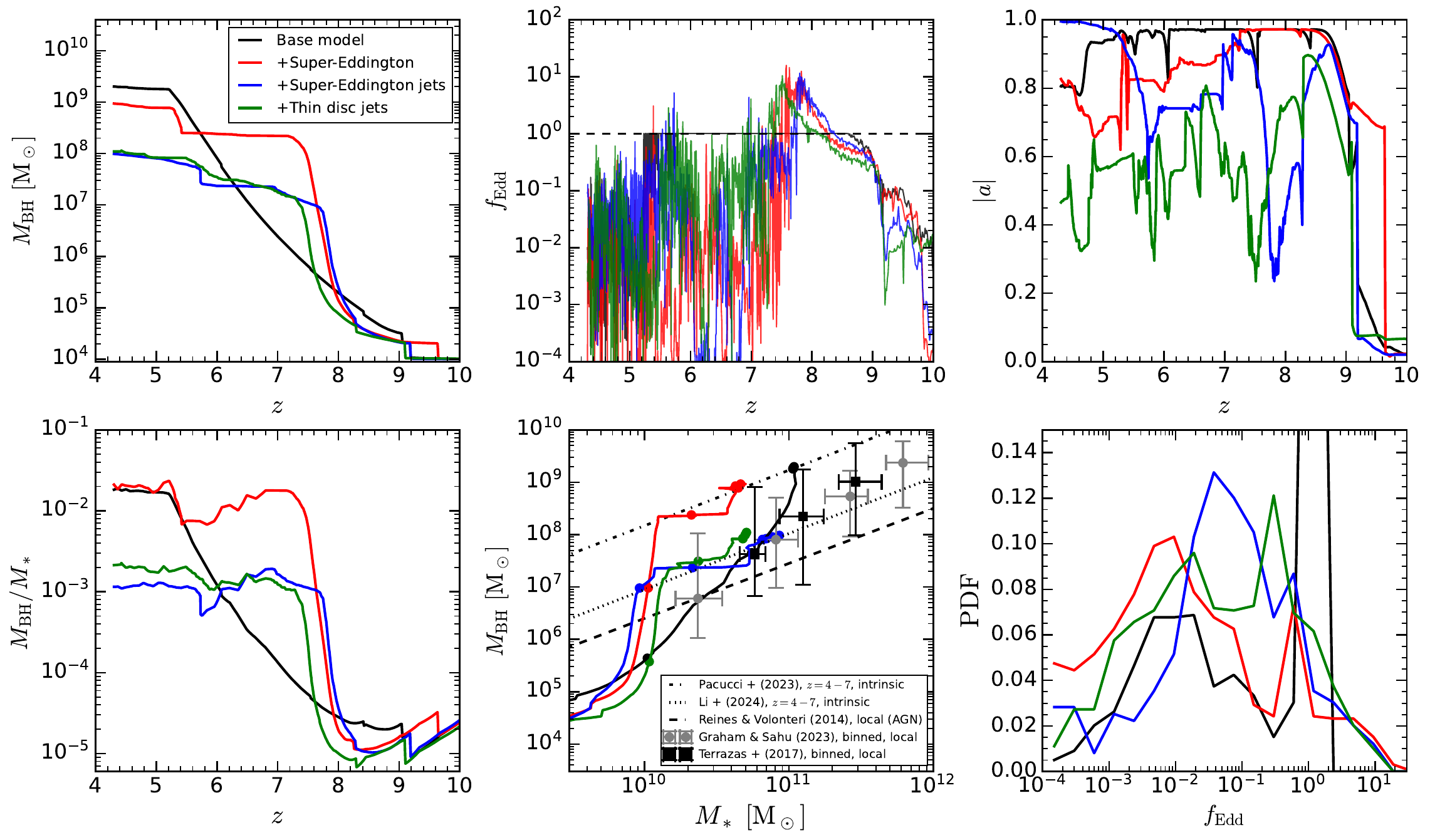}
\caption{The BH growth properties in our simulations. The panels show the evolution of the BH mass, the Eddington ratio and the BH spin magnitude in the top row from left to right, respectively. In the bottom row, the panels show the evolution of the BH mass$-$stellar mass ratio, the BH mass$-$stellar mass diagram (with the observational data described in the text) and finally the Eddington ratio probability distribution functions (ERDFs), from left to right. The ERDFs are calculated using all snapshots (1 Myr spacing beginning at $z=20$) only for the main BH in the simulations (the main progenitor of the most massive BH). The peak value of the ERDF in the 'Base model' simulation is equal to 0.37. In the bottom middle panel, the evolution curves in the BH mass$-$stellar mass diagram are marked with circles at $z=8,7,6,5,4.3$, from left to right.}
\label{fig:fig2}
\end{figure*}%

In Fig.~\ref{fig:fig2} we show the variation with redshift of several quantities related to the growth of the BHs, as well as the growth track of BHs on the BH mass - stellar mass diagram, and the Eddington ratio distribution. In the top left panel we show the variation of BH mass with redshift. The BH growth begins in earnest once the BH experiences its first merger and roughly doubles its mass from its initial seed mass ($M_\mathrm{BH,0}=10^4$ M$_\odot$). This occurs at $z\approx9$ in all of the simulations except the ‘Thin-disc-jets’ one, where it occurs slightly earlier ($z\approx9.6$). This difference is likely not of physical origin, but merely run-to-run variation due to the stochastic nature of the simulations (\citealt{Borrow2023}). 

After the initial merger, the BH in the ‘Base model’ simulation grows smoothly until it reaches a mass of $M_\mathrm{BH}=2\times10^9$ M$_\odot$ by $z=5.2$. The other three simulations, where super-Eddington accretion is allowed, all show steeper growth and overtake the ‘Base model’ case by $z=7.5$. In the ‘+Super-Eddington’ simulation, the BH grows from a very low (near-seed) mass to a value of $M_\mathrm{BH}=3\times10^8$ M$_\odot$ in a short time-span between $z=8.5$ to $z=7.5$. It then does not grow much until it experiences another burst of growth at $z=5.5$, triggered by a galaxy merger, after which it reaches $M_\mathrm{BH}=10^9$ M$_\odot$. The ‘+Super-Eddington jets’ and ‘+Thin disc jets’ simulations do not show BH masses as high as the ‘+Super-Edington’ case, at any time, reaching a maximum of $M_\mathrm{BH}=10^8$ M$_\odot$. The reduced BH growth in the simulations with jets could be for two different reasons. The first is that super-Eddington accretion, with large enough values of BH spin, features high jet efficiencies and therefore low accretion efficiencies (see Eqn.~\ref{eq:growth} in \S~\ref{sec:sec2}). In this scenario, the BH can stifle its own growth, or even reduce its mass, while releasing large amounts of energy and spinning itself down. Another possible reason for a smaller amount of BH mass growth in the simulations with jets is self-regulation: the BH has a higher feedback efficiency, and it can thus quench its own (and the galaxy's) growth given a smaller amount of accretion. It seems that in our simulations, the second of these two effects dominates (see the next subsection).

In the top middle panel we show the Eddington ratio versus redshift. The ‘Base model’ simulation features BH growth at the Eddington rate from $z\approx9$ to $z\approx5.5$, explaining the smooth growth of the BH seen in the previous panel. In all of the other three simulations, the BH growth peaks at $10-20$ times the Eddington rate, after which there is a cessation of BH growth, seen in a steep drop of the Eddington ratio down to $f_\mathrm{Edd}=10^{-3}-10^{-1}$, depending on the simulation. Thereafter the Eddington ratio is relatively low and varies strongly. At $z\approx5.5 $ there is another small and short-lived super-Eddington phase.

In the top right panel we show the evolution of the BH spin magnitude. In the ‘Base model’ case, the BH is at the spin cap $a=1$ most of the time after it starts growing significantly. This is a result of prolonged accretion in the thin disc regime, where the BH tends to spin up as long as there is a fuel supply\footnote{This spinup is efficient in the thin disc regime irrespective of the stability of the angular momentum of the gas around the BH (and how much it varies from time-step to time-step). This is owing to the fact that the realignment time-scale for spin is much shorter than either the growth time-scale of the BH, or the time-scale of evolution of the spin magnitude.}. However, short-lived dips to smaller values can be seen at specific times (e.g.~$z=8.4$, $z=7.5$, $z=6$). This coincides with small ‘jumps’ in the BH mass visible in the top left panel, indicating that this is a result of BH-BH mergers. In the other three simulations, the BH spin magnitude is, on average, lower. In the ‘+Super-Eddington’ simulation, it begins to be lower than the cap value after $z=7.5$, which coincides with the time when strong BH growth stops in this simulation. The lower BH spin magnitude is then likely due to a non-coherent gas angular momentum direction and longer alignment time-scales, despite the fact that only thin disc accretion is included in this simulation, like in the ‘Base model'. In the two simulations with jets, the BH spin is even lower and never reaches the cap value. This is a consequence of jet spindown. The BH spin drops most rapidly during the period of super-Eddington accretion, when the jet efficiency is high, as is the jet-induced spindown rate (see \S~\ref{sec:sec2}). Such behaviour is in good agreement with a recent similar study by \cite{Lupi2024}.

In the bottom left panel we show the evolution of the ratio of BH mass to stellar mass (note that we discuss the stellar mass growth in \S~\ref{res:galaxies}). In all simulations, the ratio grows during the initial Eddington or super-Eddington growth burst, until it reaches a value typical of that simulation ($\approx2\times10^{-2}$ for the two simulations without jets, and $\approx2\times10^{-3}$ for the two simulations with jets). The ratio then oscillates around that value until the end of the simulation. In the ‘Base model', the ratio grows smoothly all the way until $z=5.5$, when the BH stops growing significantly. This indicates that self-regulation is not achieved until that time.

In the bottom middle panel we show the growth tracks on the BH mass - stellar mass diagram, including the median local relations based on \cite{Graham2023} and \cite{Terrazas2017}\footnote{These two studies provide data on BH masses and stellar masses for galaxies of different types: \protect\cite{Graham2023} giving data on different morphological types (spirals, ellipticals, lenticulars) and \protect\cite{Terrazas2017} on star-forming and passive galaxies. We use this data to estimate the overall median BH mass$-$stellar mass relation for the entire population. We do this by computing weighted median BH masses in a given stellar mass bin (taking both mass errors into account), where we use morphological and passive fractions as weights. For this purpose we use morphological fractions from \cite{Moffett2016} and passive fractions from \cite{Davies2019}, both using data from the GAMA survey.} and the local AGN relation from \cite{Reines2014}, which lies somewhat below the overall median relation. We also show the high-redshift relations \text{inferred} from JWST data in two different ways: \cite{Pacucci2023} using a BH mass-limited sample ($M_\mathrm{BH}>10^{6.2}$ M$_\odot$) and \cite{Li2024} using an AGN luminosity-limited sample ($L_\mathrm{bol}>10^{44.1}$ erg~s$^{-1}$). The former relation lies roughly one to two orders of magnitude above the local median relations (depending on stellar mass), while the latter is consistent with the local median (not AGN) relations.

We find that our simulations are in broad agreement with these observed relations. The ‘Base model’ is consistent with the local median relations at all times, straddling the upper range of observed values (and agreeing with the \citealt{Pacucci2023} relation) at the end of the simulation. The ‘+Super-Eddington’ simulation straddles the upper range of the local relations and the \citealt{Pacucci2023} relation at all times (once BH growth commences). The two jet simulations are consistent with the local relation at all times. Depending on which of the two observed high-redshift relations we take to be more reliable (\citealt{Pacucci2023} or \citealt{Li2024}), our jet simulations either feature BHs with too low masses, or they are in very good agreement with the data. However, the BH masses in our jet simulations (which do not exceed $M_\mathrm{BH}=10^8$ M$_\odot$ by $z=4$) are seemingly in disagreement with the fact that there are many quasars observed at redshifts $z>6$ with masses $M_\mathrm{BH}>10^9$ M$_\odot$ (e.g.~SDSS; \citealt{Fan2001} or \textit{Chandra} survey fields; \citealt{Vito2019}). However, these latter observations cover a much larger comoving volume than we are simulating here, since they cover much of the sky. While our simulations are of a protocluster (therefore a rare environment with a BH that starts growing fairly early), it is only the sixth most massive system (at $z=4.3$) in a 400 cMpc volume. Note that such a comoving volume corresponds to survey areas of $\approx3$ deg$^2$ for redshift intervals of $z=4-6$, $z=6-8$ and $z=8-10$. This is much smaller than the area covered by SDSS, so observations can find even rarer objects with even earlier BH growth, reaching even larger BH masses at these high redshifts. The observed quasar samples are also likely heavily biased towards brighter and more massive BHs (see \citealt{Tanaka2024} and \citealt{Li2024} for discussions of the impact of selection effects on AGN observability).

Finally, in the bottom right panel we show the Eddington ratio probability distribution functions (PDFs) for the different simulations, using all snapshots of the simulation (which are uniformly spaced, 5 Myr apart). The ‘Base model’ shows a large peak at $f_\mathrm{Edd}=1$, because the BH spends most of the time at that value, as well as a small one at $f_\mathrm{Edd}=10^{-2}$. The other simulations show one peak at or near the Eddington value, as well as (in some of the simulations) another one near $f_\mathrm{Edd}=10^{-2}$. There is a long tail of values extending to and beyond $f_\mathrm{Edd}=30$.

Overall, the results from Fig.~\ref{fig:fig2} indicate that Super-Eddington accretion should be allowed in simulations of galaxy formation. Otherwise, the BH can grow at exactly the Eddington rate for very long times, which is entirely artificial behaviour. In this case, the BH takes a very long time to reach a self-regulated phase of growth, and thus an equally long time to start affecting its galaxy (see further discussion in \S~\ref{res:galaxies}). Furthermore, it is not clear that sufficiently high values of BH masses can be achieved sufficiently early to match observations, although this is also not clearly excluded by these simulations. Our results here are in good agreement with a similar study by \cite{Bennett2024}, who compared the fiducial and super-Eddington cases of a simulated protocluster using the FABLE model.

\begin{figure*}
\includegraphics[width=\textwidth, trim = 0 10 0 0]{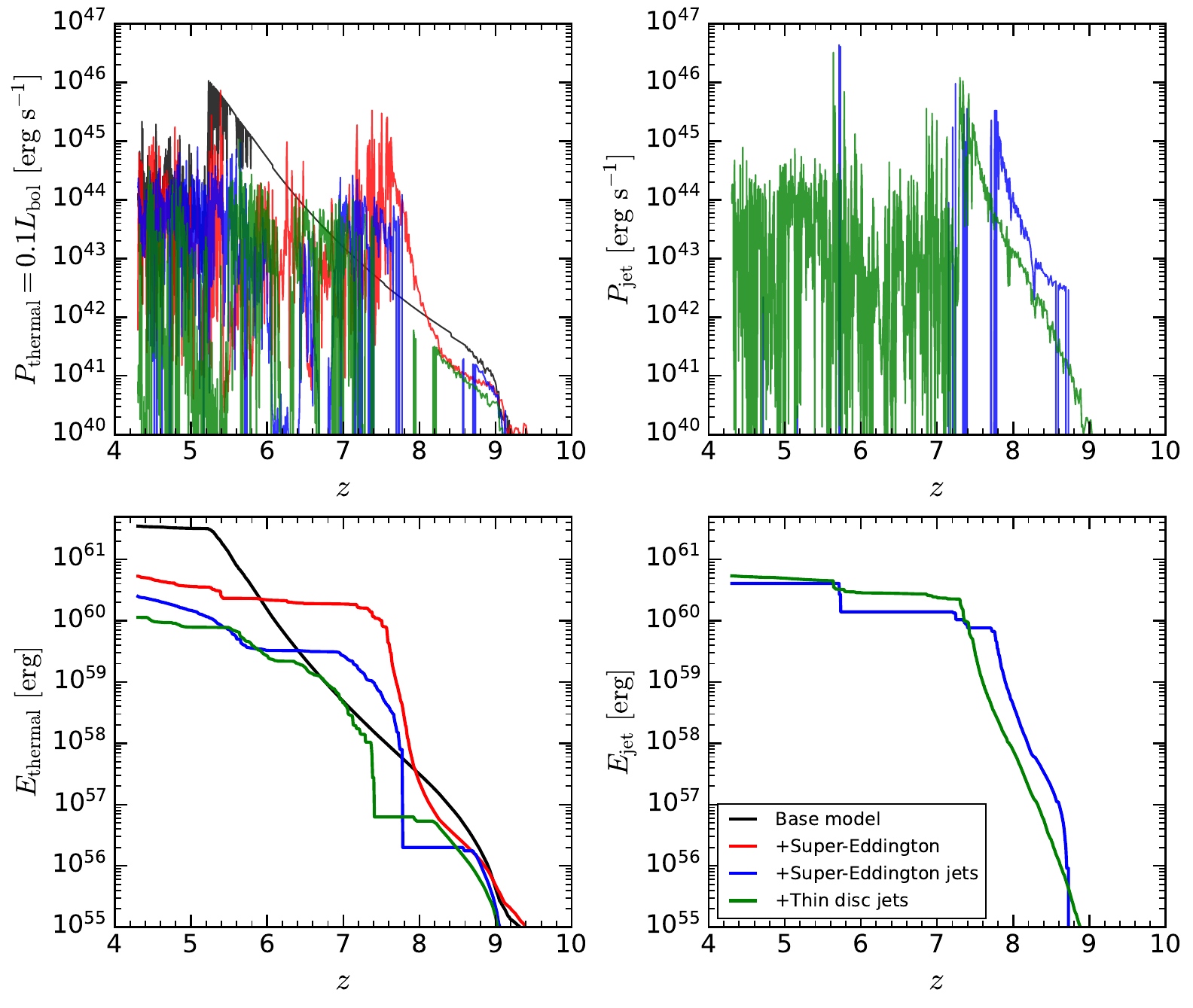}
\caption{The evolution of the feedback properties. In the top row we show the thermal feedback (which is 10 per cent of the bolometric luminosity for the thin disc) and jet powers, while in the bottom row we show the cumulative injected thermal and jet energies.}
\label{fig:fig3}
\end{figure*}%

The comparisons shown here also indicate that simulations based on $\Lambda$CDM are not in nearly as strong disagreement with observed BH masses as claimed by some recent studies. This is despite the fact that we have used a relatively ‘generic’ galaxy formation model (the EAGLE model) that is calibrated to $z=0$ data and is by no means designed to reproduce such high redshift data. On the other hand, in this paper we have focused on a protocluster, in which both BH and galaxy growth starts earlier, and in which the BH may tend to have higher masses, compared to the overall BH population, at a given stellar mass at that redshift.

In the case that the high-redshift BH mass$-$stellar mass relation inferred by \cite{Pacucci2023}, from JWST data, is truly representative of the intrinsic relation at that redshift, our jet simulations likely feature insufficiently high BH masses, as a consequence of high jet efficiencies and additional BH self-regulation. This could possibly be remedied by suppressing the accretion rate in the super-Eddington regime (due to the effects of accretion-driven disc winds that take away most of the accreting mass). In this case, the BH growth would be suppressed in the super-Eddington phase, but so would its feedback. This could possibly lead to longer-lived high-accretion phases, but it is not clear which effect on the BH growth would win out (the explicit suppression, or the subsequent boost from longer-lived accretion episodes).

\subsection{Black hole feedback}
\label{res:bh_feedback}

\begin{figure}
\includegraphics[width=\columnwidth, trim = 0 10 0 0]{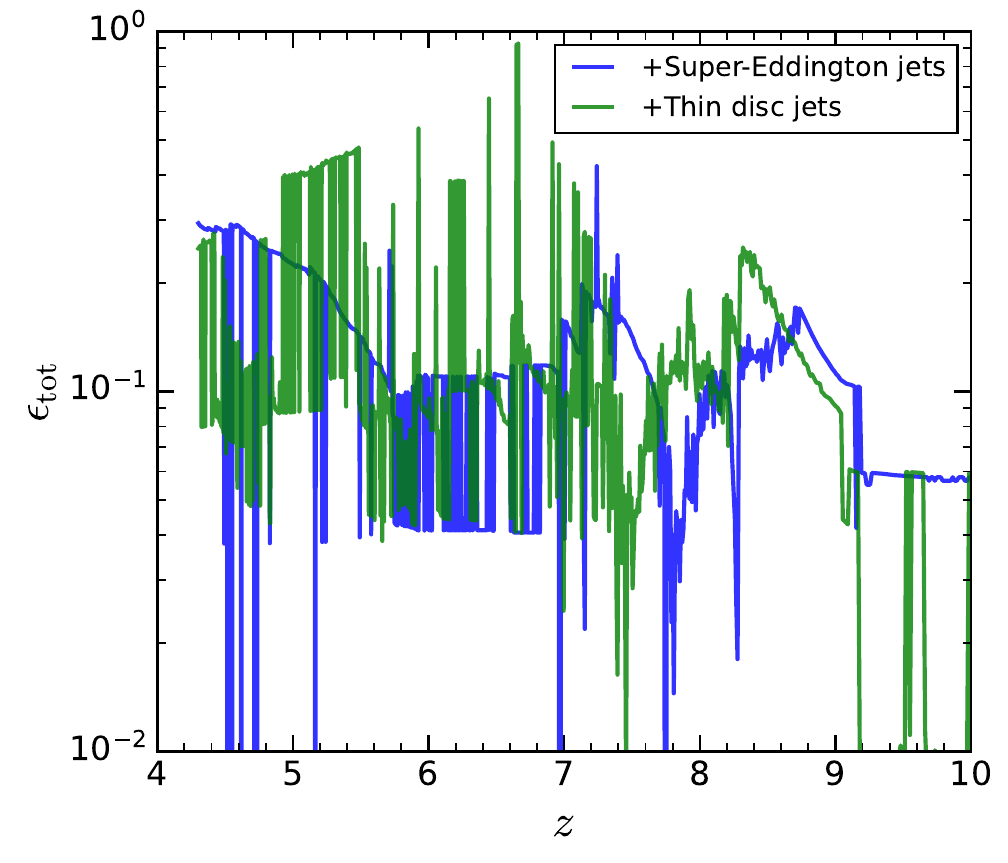}
\caption{The total energy release efficiency (the sum of the jet, radiative and wind efficiencies) in our two jet simulations. The energy release efficiency is equal to 10 per cent for both the 'Base model' and '+Super-Eddington' simulations, which are not shown here. The efficiencies are computed as explained in \S~\ref{sec:sec2}.}
\label{fig:fig34}
\end{figure}%

We now proceed to discuss the feedback effects of the BH in the different simulations. In Fig.~\ref{fig:fig3} we show the evolution of the thermal and jet feedback powers (top row), as well as the total injected energies in the two channels (bottom row). The top left panel shows that the thermal feedback power, which is equal to 10 per cent of the bolometric luminosity in our simulations for the thin disc, in the ‘Base model’ is limited by the BH mass and reaches significant values only at fairly late redshifts (values of e.g.~$P_\mr{th}=10^{44}$, $10^{45}$ and $10^{46}$ erg~s$^{-1}$ are reached only at $z=6.5$, $z=6$ and $z=5$, respectively). On the other hand, the ‘+Super-Eddington’ run has a significant thermal outburst ($P_\mr{th}=10^{45}$ erg~s$^{-1}$) already at $z=7.5$ (coinciding with the strongly super-Eddington accretion episode visible in Fig.~\ref{fig:fig2}. The two jet runs do not have significant thermal outbursts (the values never exceed $P_\mr{th}=10^{45}$ erg~s$^{-1}$ and rarely tip over $P_\mr{th}=10^{44}$ erg~s$^{-1}$ in both the simulations). In the top right panel, we show the evolution of the jet power with time for the two jet simulations. The largest values of the jet power ($P_\mr{jet}\approx10^{46}$ erg~s$^{-1}$) are achieved near $z=7.5$ and $z=5.5$, when the BH becomes super-Eddington (as a consequence of the strong growth of the jet efficiency with Eddington ratio, see \S~\ref{sec:feedback_effs_jet}).

From the bottom left panel, we see that the ‘Base model’ and ‘+Super-Eddington’ simulations feature the largest injection of thermal energy. The ‘+Super-Eddington’ features the largest amount of injection at higher redshifts (e.g.~$z=7$), but the ‘Base model’ eventually catches up and has the largest injected amount by $z=5$ (in agreement with the fact that it features a large thermal feedback power, $P_\mr{th}>10^{45}$ erg~s$^{-1}$, for a prolonged time between $z=6$ and $z=5$, while the ‘+Super-Eddington’ simulation only shows such high values in shorter bursts). Both of the jet simulations have injected a few times less thermal energy than ‘+Super-Eddington', and more than an order of magnitude less than ‘Base model’ by $z=4.5$. Finally, in the bottom right panel we show the injected jet energies. By $z=7$ (the redshift by which the initial super-Eddington episode finishes in the three super-Eddington simulations), the BHs in the jet simulations have injected an almost identical amount of jet energy, $E_\mathrm{jet}\approx3\times10^{60}$ erg, which is also very similar to the thermal energy injected in the ‘+Super-Eddington’ simulation by that redshift. This could be a coincidence, but it is more likely that this is the feedback energy that needs to be injected at this redshift to initiate self-regulation. 

In Fig.~\ref{fig:fig34} we show the evolution of the total energy release efficiency, which is the sum of the jet, wind and radiative efficiencies (determined as explained in \S~\ref{sec:sec2}, see also Fig.~\ref{fig:fig0} and its caption). We show these only for the two runs with jets, since for both the 'Base model' and '+Super-Eddington' simulations, the total energy release efficiency is simply the radiative efficiency, which is 10 per cent. We find that in both of the jet simulations, the feedback efficiency varies around a similar value as in the two thermal-only runs (10 per cent), but it can be much more intermittent. This is due to the sudden changes in the efficiencies when accretion/feedback regimes change at the Eddington rate. For the run with jets included in the thin disc regime, at sub-Eddington accretion rates, the feedback efficiency is on average higher and also more intermittent, both of which is caused by the inclusion of the jet component.

The total efficiency sometimes reaches relatively large values (of order 50 per cent or more), but this is quite rare and very-short lived (note that it is also more frequent in the '+Thin disc jets' simulations). Both of the jet simulations show the stifling of their BHs' growth by $z=7.5$ (see top left panel in Fig.~\ref{fig:fig2}), after which point the '+Super-Eddington' simulation instead shows continued BH growth. The stifling of the BH growth in the jet simulations occurs at a time when their total energy release efficiency is modest (no more than 20-30 per cent). This indicates that the direct impact of energy release, in the form of the $(1-\epsilon_\mathrm{tot}$) accretion efficiency factor present in Eqn.~\ref{eq:growth}, is not the dominant factor that causes the stifling of BH growth. We instead prefer the interpretation laid out earlier in this subsection; we suspect the BH growth is stifled as a result of self-regulation. This picture is supported by the BHs injecting similar amounts of energy in different simulations, regardless of different feedback modes or efficiencies that were adopted.

\subsection{Impact on large scales}
\label{res:bh_feedback_large_scale}

\begin{figure*}
\includegraphics[width=\textwidth, trim = 0 12 0 0]{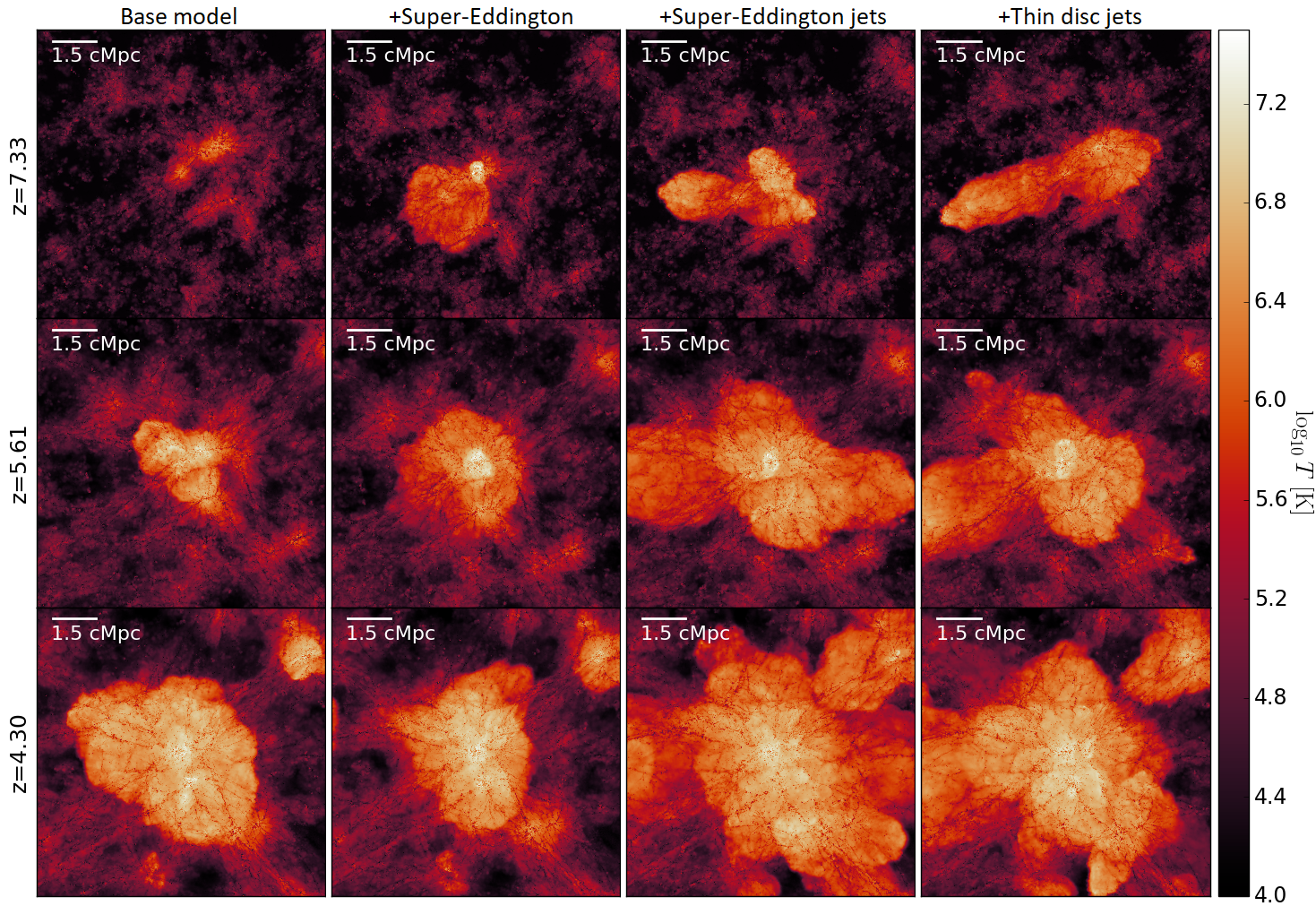}
\caption{The impact of BH feedback on the forming protocluster, as shown here through mass-weighted temperature projections (10 Mpc deep, equal to the side-length of the images). Different columns show different simulations, while rows show different redshifts, both as labeled. }
\label{fig:fig4}
\end{figure*}%

In Fig.~\ref{fig:fig4} we show the impact of feedback in the four simulations (different columns) on the large-scale surroundings of the BH and galaxy as the protocluster is forming. We show temperature projections (equally deep as the images are large) at three different redshifts, as labeled. At $z=7.33$ (around the time of the first and strong super-Eddington episode), we see that the ‘Base model’ has no significant consequences of feedback yet visible, while the other three simulations all show signs of outflows. In the ‘+Super-Eddington’ simulation these are spherical, while in the remaining two (with jets), they are more directed. At the next redshift ($z=5.61$), all four runs show some signs of outflows, but they are clearly stronger in the two jet cases. At the final redshift $z=4.31$, all four runs shown significant impact on the surroundings, with the jet cases still showing the most dramatic effects. Interestingly, the ‘Base model’ shows stronger outflows than the ‘+Super-Eddington’ simulation, likely due to the BH catching up in its growth and injecting more energy (see the previous subsection).

\begin{center}
\begin{figure*}
\includegraphics[width=0.9\textwidth, trim = 0 13 0 0]{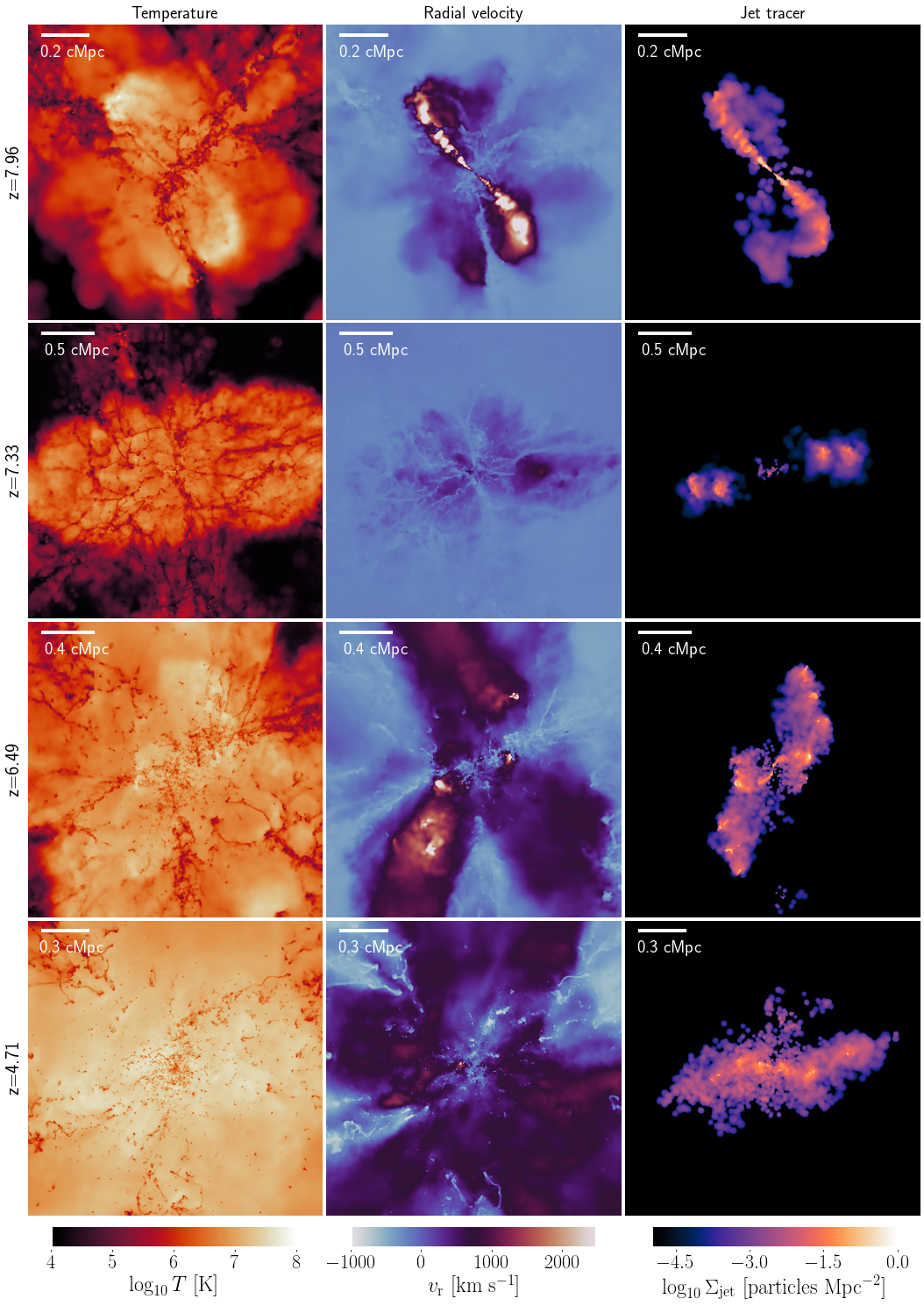}
\caption{Jet activity in the ‘+Super-Eddington jets’ simulation. We show the temperature, radial velocity (with respect to the central BH) and jet tracer (surface density of particles launched into jets) projections at different times, as labeled. The projections are equally deep as the maps are large (the sizes of the maps vary with redshift).}
\label{fig:fig5}
\end{figure*}%
\end{center}

In Fig.~\ref{fig:fig4}, actual jets are not visible as such (only the outflows, shocks and bubbles they induce in the ambient medium). For this reason, we show the direct effect of jets in Fig.~\ref{fig:fig5} through a few different quantities, using the ‘+Super-Eddington jets’ simulation. At several different times and in different columns we show projections of the temperature, radial velocity (with respect to the central BH) and the jet tracer (the surface number density of particles that have been launched into jets). At all times we can see cosmological infall and accretion of gas through cold gas filaments. Outflows emanate in all other directions simultaneous to the inflow. At late times, the outflows are strong enough to disrupt the filamentary inflow. The jets are usually not directly visible in the temperature projections, but their effects through shock heating are visible. This is especially true at the earliest time $z=7.96$, when heating at the jet hotspots is visible, and at $z=7.33$, when an ellipsoidal bow shock is clear. The jet tracer shows a clear continuous jet at the first time, while at later times, episodic activity can be inferred from the gaps and density peaks present in the jet tracer maps.

\subsection{Impact on galaxy growth and properties}
\label{res:galaxies}

\begin{figure*}
\includegraphics[width=0.98\textwidth, trim = 0 20 0 0]{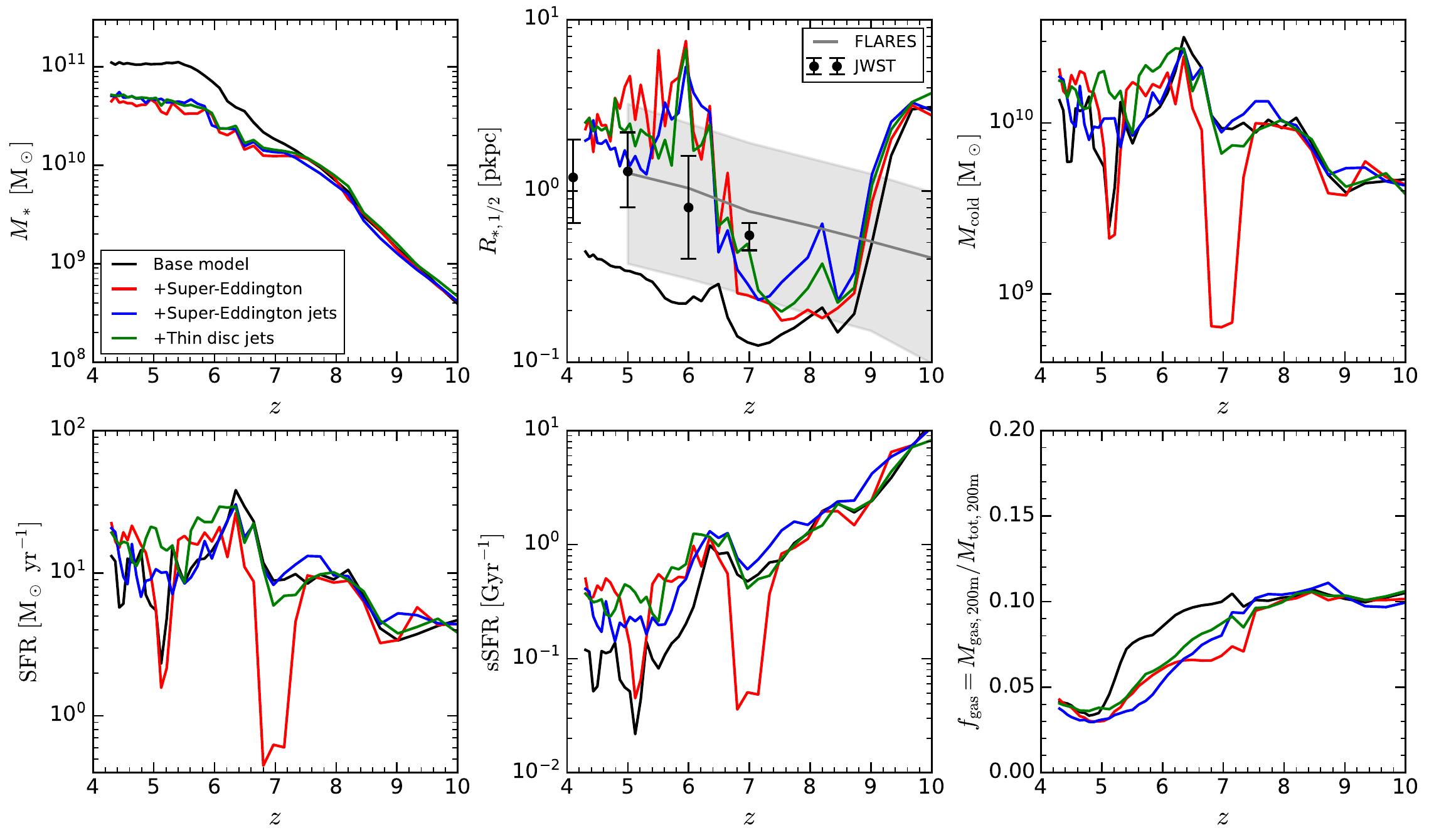}
\caption{The evolution of main galaxy-related properties. We show, in order from left to right and top to bottom, the: 1) stellar mass, 2) stellar half-mass radius, 3) cold gas mass ($T<3\times10^4$ K), 4) star formation rate, 5) specific star formation rate, 6) total gas fraction in the DM halo (including hot and cold gas). The galaxy size data from FLARES and JWST, shown in the top middle panel, are from \protect\cite{Roper2022} and \protect\cite{Chworowsky2024}, respectively. The latter is for all galaxies with $M_*>10^{10}$ M$_\odot$, and defined as the effective radius. The simulated galaxy sizes are not reliable until $z\approx8$ due to miscentering between the BH and the galaxy.}
\label{fig:fig6}
\end{figure*}%

Finally, we discuss the impact of the BH's feedback on its host galaxy. In Fig.~\ref{fig:fig6} we show the evolution of a set of main galaxy properties of interest. These are all computed in apertures of 30 pkpc. We begin with the stellar mass, shown in the top left panel. In the ‘Base model', the stellar mass of the galaxy grows continuously until $z=5.5$, by which point it has reached a value of $M_*=10^{11}$ M$_\odot$. At this redshift, the stellar mass growth is halted, which is likely due to the BH growth catching up and causing a large feedback event (Figs.~\ref{fig:fig2} and \ref{fig:fig3}). In the other three simulations, the stellar masses are very similar to each other, but smaller than in the ‘Base model'. The growth is already halted at $z=7.5$ (coinciding with the feedback event at that redshift visible in Fig.~\ref{fig:fig3}). It continues again at $z\approx6$, but is again slowed down by $z\approx5.5$ and continues slowly by the end of the simulations at $z=4.3$. The stellar mass at this final redshift is only half as large in these three simulations as it is in the ‘Base model', showing that turning on super-Eddington accretion can significantly affect fundamental galaxy properties. The two jet simulations appear to have slightly but systemically larger stellar masses than the ‘+Super-Eddington’ run at most redshifts, likely caused by the fact that the ‘+Super-Eddington’ run has a more massive BH (see Fig.~\ref{fig:fig2}), which may be more efficient at quenching the galaxy, even with a lower feedback efficiency.

In the top middle panel, we show the galaxy's stellar half-mass radius. In all simulations, the galaxy is a few kpc in radius at very high redshifts ($z\approx10$). However, we center on the BH in our simulations, and the BH is experiencing mergers at this redshift and is thus not centered perfectly on the galaxy. The stellar half-mass radius is thus overestimated at these very high redshifts. By $z\approx8$ the BH is growing in a dense core of gas, which is both highly star-forming and starting to feed the BH. In the 'Base model', the galaxy remains small until it reaches a minimum of $\approx100$ pc around $z=7$, despite its BH growing at the Eddington rate from $z=9$, since it is not massive enough to halt the galaxy's mass growth. Beyond $z=7$, the galaxy's half mass radius begins to increase. This is known to be a common result of stronger AGN feedback (e.g.~\citealt{Sales2010}). In our case, it is not clear which physical mechanism is responsible, but we speculate it could a result of either 1) preferential ejection of low-angular-momentum gas, and subsequent formation of stars at larger radii, or 2) migration of stars that formed in the core as an adiabatic response to changes in the gravitational potential (due to gas outflows caused by feedback) (see e.g.~\citealt{Choi2018} and \citealt{Vlugt2019} for a discussion of both mechanisms). In the 'Base model', the galaxy reduces in size again around $z=6$ due to a growth burst from a high SFR (which is concentrated in the dense gas core), before continuing to expand again. However, by the end of the simulation, the galaxy is only $0.5$ kpc in radius.

The super-Eddington runs, on the other hand, reach a minimum half-mass radius already around $z=8.5$, after which point the galaxies start growing and reach larger sizes than in the 'Base model'. This difference in sizes is due to BHs in the super-Eddington runs halting the galaxies' mass growth earlier in time, allowing the positive size evolution from adiabatic expansion to commence earlier and to result in larger sizes by the end of the simulation. The sizes reach values of a few kpc in the super-Eddington runs by $z=6$ and then remain similar to those values until the end of the simulation. We discuss the detailed comparison with JWST and FLARES galaxy sizes in the next subsection.

In the top right panel, we show the evolution of the cold gas mass (defined as all gas with $T<3\times10^4$ K). In all cases there is a reservoir of $M_\mathrm{cold}\approx10^{10}$ M$_\odot$ feeding the galaxy by $z=8$. The feedback event at $z=7.5$ is only visible here in the ‘+Super-Eddington’ run in the form of a large dip in the cold gas mass. By $z=6.5$, however, the gas mass again increases to match the other simulations, at $M_\mathrm{cold}\approx2\times10^{10}$ M$_\odot$. After this, the cold gas mass stays stable around that value or shows a slight decrease. In the two simulations without jets, however, there is a further sharp drop just after $z=5.5$ (corresponding with the other super-Eddington episode), which could indicate that gas can more easily be expelled by isotropic thermal outflows rather than directed kinetic jets.

In the bottom-left and middle panels, we show the star formation rate (SFR) and specific SFR (sSFR), respectively. These do not differ significantly between the models, with the exception of the ‘+Super-Eddington’ runs featuring a significant drop in the SFR/sSFR at $z\approx7$, which is likely caused by a strong, isotropic feedback event that disrupts filamentary accretion onto the central galaxy. If this galaxy were observed during this period, it would be considered quiescent. However, even in that run, the SFR quickly recovers, likely due to filamentary accretion onto the central galaxy recommencing. The galaxy is thus only transiently quiescent. The other runs do not experience such an SFR/sSFR drop, probably due to their filamentary accretion never ceasing. 

The values of the SFR/sSFR shown in these panels at first do not seem consistent with the top-left panel, where it is clear that in the ‘Base model’ there is much more growth between $z=7$ and $z=5.5$ than in the other models, despite the SFRs being similar. Integrating the SFRs, we find that the galaxy increases its stellar mass by $2-3\times10^{10}$ M$_\odot$ between $z=8$ and $z=4.3$ in the different simulations. This is only half of the galaxy stellar mass by $z=4.3$ in the three models other than ‘Base model', and it is only a quarter of the final stellar mass of the galaxy in that model. It thus appears that allowing super-Eddington accretion reduces the growth of the galaxy largely by changing the properties/SFRs in satellite galaxies (both within and outside the apertures that we use to compute galaxy properties) that are merging onto the primary, rather than by directly affecting the primary.

Finally, in the bottom right panel we show the evolution of the total gas fraction (including both hot and cold gas) in the DM halo. The gas fraction is similar in all four simulations and stable at $10$ per cent (compare to the cosmological value of $15$ per cent) until $z=7.5$, after which it begins to drop in all simulations except ‘Base model’ owing to significant feedback. In the ‘Base model', this reduction begins slightly later and is more smooth. By the end, all four simulations reach a very similar gas fraction of around $4$ per cent.

\subsection{Comparison with JWST observations}
\label{res:JWST}

We will now discuss the results shown in Fig.~\ref{fig:fig6} in the context of recent JWST observations at high redshift. We caution the reader to bear in mind that the stellar masses and SFRs of observed sources are based on spectral energy distribution (SED) fitting, which is model-dependent, and also on details of the data reduction and source extraction pipelines. Furthermore, the observed sizes of galaxies depend on the surface brightness limit of the survey, and are defined as the effective radius, i.e. the projected radius within which half of the (observed) galaxy's light is emitted. We also remind the reader that we selected the 6th most massive protocluster in a (400 cMpc)$^3$ cube; as such it is a rare system but not representative of the more extreme ones which could be found in larger parent volumes.

\cite{Weibel2024} recently found a massive, quenched and concentrated galaxy at $z=7.3$. The galaxy has a stellar mass of $M_*=10^{10.2}$ M$_\odot$, specific star formation rate of $\mr{sSFR}<10^{-1}$ Gyr$^{-1}$ and effective radius of $r_\mr{e}=200$ pc. This stellar mass is slightly higher than the stellar masses of our galaxies at the same redshift ($M_*=10^{10.0}$ M$_\odot$). The sSFR is larger in all of our galaxies at the same redshift, except in the ‘+Super-Eddington simulation'. The half-mass radius of our galaxies is very similar to the effective radius of the observed galaxy in all three of our super-Eddington simulations, while it is somewhat too small in our ‘Base model’ (120 pc), compared to this galaxy. These comparisons indicate that the galaxy observed by \cite{Weibel2024} may be the central galaxy of a protocluster similar to the one we are simulating. 

The size evolution of our galaxy in the different simulations can also be compared with the data recently published by \cite{Chworowsky2024} for a sizeable sample ($\approx100$) of galaxies with $M_*>10^{10}$ M$_\odot$ at high redshift ($z>4$). In Fig.~\ref{fig:fig6} (top middle panel) we show the median and 16th to 84th percentile range of their observed galaxy sizes at $z=4,5,6,7$. Our galaxy in the runs that allow super-Eddington accretion agree with the median sizes at $z=7$, and are slightly larger than the median size at $z<6$ (but within $2\sigma$ of their sample). The ‘Base model', on the other hand, always features much smaller galaxy sizes, albeit also within $2\sigma$ of the observed sample. For reference, we also include the median galaxy sizes from the FLARES simulations in the relevant redshift range (\citealt{Roper2022}). Note that our simulations are not expected to perfectly agree with the FLARES results since FLARES used the EAGLE galaxy formation model (\citealt{Schaye2015}), while we use the SWIFT-EAGLE model (\citealt{Schaller2024}), which is different in many minor aspects, mostly numerical in nature.

\section{Summary and conclusions}
\label{sec:conclusions}

Using the SWIFT code and an updated version of the EAGLE galaxy formation model, we have simulated the evolution of a protocluster down to redshift $z=4.3$ using the zoom-in method. At this redshift, the primary halo has a mass of $10^{13}$ M$_\odot$. We performed several simulations, focusing on the black hole physics and varying assumptions about black hole accretion and feedback. The progenitor of the primary black hole in the simulations is seeded with a mass of $10^4$ M$_\odot$ at redshift $z\approx17$, and it thereafter grows through Bondi-Hoyle-Lyttleton accretion (without any boost or suppression factors), as well as mergers.

We compare the ‘Base model', which uses an Eddington limit on the accretion rate and thermal isotropic feedback, with one where super-Eddington accretion is allowed (but without any further modifications), as well as runs where accretion physics is also included to model BH spin and launch AGN jets, either only at super-Eddington accretion rates (the third simulation) or also at sub-Eddington rates (the final simulation).

From studying the BH growth (Fig.~\ref{fig:fig2}) and feedback (Fig.~\ref{fig:fig3} and~\ref{fig:fig34}) in these simulations we find the following:
\begin{itemize}
    \item If super-Eddington accretion is not allowed, the BH grows without interruption at the Eddington rate for a long time, from $z=9$ to $z=5.5$. On the other hand, if super-Eddington accretion is allowed, the BH grows by a very large amount in a short time, reaching Eddington ratios of $f_\mathrm{Edd}\approx20$ and quenching its own growth by $z=7$. The BH goes through another super-Eddington phase around $z=5.5$.
    \item The final BH mass at $z=4.3$ is largest in the ‘Base model’ ($M_\mathrm{BH}=2\times10^9$ M$_\odot$), and second largest in the ‘+Super-Eddington’ run without jets ($M_\mathrm{BH}=10^9$ M$_\odot$). This is likely because the BH ‘overshoots’ in the ‘Base model', having done less feedback at earlier times. The two jet runs both feature relatively low final BH masses ($M_\mathrm{BH}=10^8$ M$_\odot$), as a consequence of higher feedback efficiencies.
    \item The BH masses are consistent with those inferred from JWST observations in our runs. Depending on how the intrinsic BH mass$-$stellar mass relations are inferred from observations, our jet runs may or may not feature too low BH masses, at a given stellar mass. If they do, accretion may need to be suppressed (using an accretion efficiency parameter that is $<100$ per cent), which could help dampen the strong feedback and self-regulation effects from jets.
\end{itemize}

Focusing then on the impact of the BH on its large-scale surroundings (Figs.~\ref{fig:fig4} and \ref{fig:fig5}) and the host galaxy (Fig.~\ref{fig:fig6}), we find that:
\begin{itemize}
    \item Allowing super-Eddington accretion has a large effect on host galaxy properties. In the ‘Base model', the galaxy grows continuously until it is quenched at a late time ($z=5.5$), while in the super-Eddington runs, it is already quenched earlier ($z=7.5$) for a short time as a consequence of the strong feedback episode, before continuing to grow at a slow pace. The final galaxy stellar mass in the ‘Base model’ is $M_*=10^{11}$ M$_\odot$, compared to $M_*=5\times10^{10}$ M$_\odot$ in the other runs. The galaxy is much more compact in the ‘Base model’ (hundreds of pc compared to a few kpc).
    \item Comparing the three simulations that allow super-Eddington accretion, but vary the implementation of AGN feedback, we find that galaxy properties are nearly unaffected. It appears that allowing super-Eddington accretion is the dominant change that can be made when it comes to impacting the properties of galaxies.
    \item The large-scale impact of the feedback in the different simulations is also fairly similar (with the exception of the ‘Base model', which only shows large-scale outflows at late times), with the only difference being that jet-driven outflows are slightly less isotropic.
\end{itemize}

The simulations and results we have presented here indicate that allowing super-Eddington accretion in cosmological hydrodynamical simulations would likely lead to improvements in statistical predictions and comparisons with observations of BH growth at high redshift, at least for protoclusters, and we thus encourage including this for future models of galaxy formation.

\section*{Acknowledgements}
FH thanks Adrian Jenkins, Yannick Bahé, Robert Crain and Azadeh Fattahi for their contributions to making this project possible. The research in this paper made use of the SWIFT open-source simulation code (\url{http://www.swiftsim.com}, \citealt{Schaller2024})
version 0.9.0. The swiftsimio Python library was used to analyze and visualize the data from the simulations (\citealt{Borrow2020_swiftsimio}, \citealt{Borrow_2021_swiftsimio}). FH would like to acknowledge support from the Science Technology Facilities Council through a CDT studentship (ST/P006744/1), and CGL acknowledges support from STFC consolidated grants ST/T000244/1 and ST/X001075/1. WJR acknowledges support from the Sussex Astronomy Centre STFC Consolidated Grant
(ST/X001040/1). This project has received funding from the Netherlands Organization for Scientific Research (NWO) through research programme Athena 184.034.002. This work used the DiRAC@Durham facility managed by the Institute for Computational Cosmology on behalf of the STFC DiRAC HPC Facility (www.dirac.ac.uk). The equipment was funded by BEIS capital funding via STFC capital grants ST/K00042X/1, ST/P002293/1, ST/R002371/1 and ST/S002502/1, Durham University and STFC operations grant ST/R000832/1. DiRAC is part of the National e-Infrastructure.

\section*{Data availability}

The data underlying this article will be provided upon reasonable request to the corresponding author. The code and initial conditions used to generate the data are available online: \url{https://gitlab.cosma.dur.ac.uk/swift/swiftsim}.

\bibliographystyle{mnras}
\bibliography{jet_bibliography} 

\appendix
\section{Slim disc wind efficiency formula}
\label{app}

\begin{figure*}
\includegraphics[width=1\textwidth, trim = 0 10 0 0]{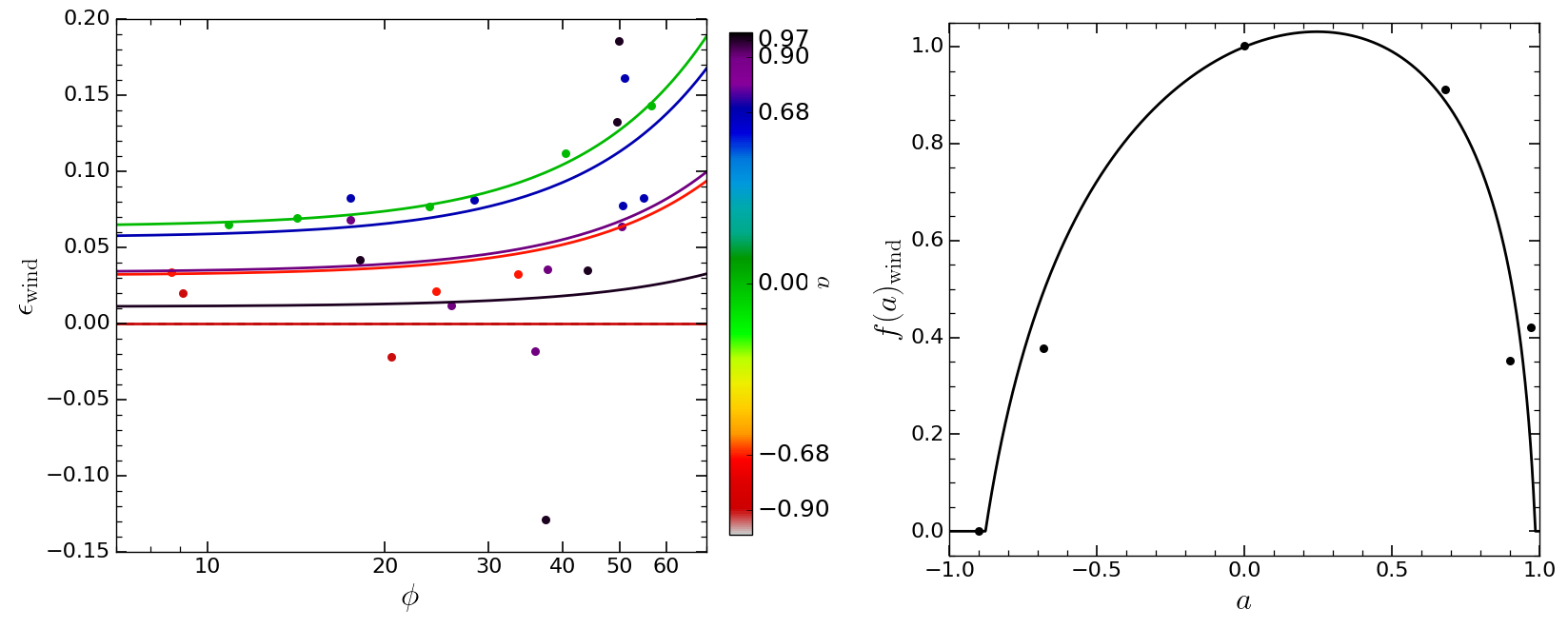}
\caption{The slim disc wind efficiencies used in our model (left-hand panel), obtained using data from the simulations performed by \protect\cite{Ricarte2023}. The points show average values of the wind efficiency for each individual simulation performed by \protect\cite{Ricarte2023}, where the $x-$axis values are the average dimensionless magnetic fluxes $\phi_\mr{BH}$ for each of those simulations. The wind efficiencies are obtained as described in the text. The lines represent the fitting function given by Eqn.~(\ref{eq:eps_wind_slim_fit}), which includes an explicit dependence on magnetic flux, which is most apparent for the zero spin case. The right-hand panel shows the remaining dependence on BH spin, $f(a)$, which we fit using Eqn.~(\ref{eq:f_a_fit}) (see text for how we obtain the individual values of that function).}
\label{fig:slim_disc_eff}
\end{figure*}%

The GRRMHD simulations performed by \cite{Ricarte2023} span a wide range of accretion rates (sub- and super-Eddington) as well as multiple values of spin. The accretion rates are variable, but they vary around some mean value, which is effectively an input for their simulations. The authors provide data on the average dimensionless magnetic flux threading the horizon, $\phi_\mr{BH}$, which reaches some saturated value if the system has reached the MAD limit (in which case the saturated value of $\phi_\mr{BH}$ depends on BH spin $a$ and accretion rate, or Eddington ratio $f_\mathrm{Edd}$).

For each simulation, the authors also provide an average value of the MHD efficiency $\epsilon_\mr{tot}$, measured at $5R_\mr{G}$. This efficiency should represent the output in both jets and winds, although it may be an underestimate of the latter since winds may experience some acceleration at large radii. Using the available data, we find the value of the wind efficiency, not provided by the authors, for each simulation by subtracting from their MHD efficiency the expected value of the jet efficiency, given by Eqn.~(\ref{eq:epsilon_jet}) from \cite{Tchekhovskoy2010}, which uniquely determines the jet efficiency for any MAD disc if one knows the BH spin and magnetic flux threading the BH horizon.

The average wind efficiencies for the slim disc, given by this procedure, are shown for each simulation performed by \cite{Ricarte2023} as points in the left-hand panel of Fig.~\ref{fig:slim_disc_eff}. For $a=0$, the efficiency clearly depends on magnetic flux in a monotonic way. We find that it can be described with the following fitting function (compare the green line and the green data points in the figure):
\begin{equation}
\epsilon_\mr{wind}(\phi_\mr{BH},a=0) = 0.065\bigg[1+\bigg(\frac{\phi_\mr{BH}}{50}\bigg)^2\bigg].
\label{eq:eps_wind_slim_zero_spin}
\end{equation}
For non-zero values of BH spin, the wind efficiencies from the simulations (the individual data points) seem to have a much larger scatter, and do not follow an obvious monotonic relation. For simplicity, we have assumed that the wind efficiency can be parameterized in the following way:
\begin{equation}
\epsilon_\mr{wind}(\phi_\mr{BH},a) = 0.065\bigg[1+\bigg(\frac{\phi_\mr{BH}}{50}\bigg)^2\bigg]f(a),
\label{eq:eps_wind_slim_fit}
\end{equation}
where $f(a)$ is a function that depends only on spin. In other words, we have assumed that the dependencies on magnetic flux $\phi_\mr{BH}$ and BH spin $a$ are decoupled (which does not necessarily have to be the case). Given this assumption, we find the average value of $f(a)$ for different values of spin $a$ by dividing $\epsilon_\mr{wind}(\phi_\mr{BH},a)$ with the fitting function $\epsilon_\mr{wind,slim}(\phi_\mr{BH},a=0)$ (given by Eqn.~\ref{eq:eps_wind_slim_zero_spin}) for each individual simulation, and then taking an average over multiple simulations (and therefore multiple values of $\phi_\mr{BH}$). This yields the functional form $f(a)$, which we show in the right-hand panel of Fig.~\ref{fig:slim_disc_eff}. We also show a simple fit to the obtained values of the function, given by
\begin{equation}
f(a) = \max(0,1+\Omega_\mr{H}-8\Omega_\mr{H}^2),
\label{eq:f_a_fit}
\end{equation}
where we have opted to parametrize the fit using the (dimensionless) angular velocity of the BH horizon, $\Omega_\mr{H}$, rather than the BH spin itself, since the former is a more physically meaningful quantity in the context of energy extraction from the BH.

\bsp	
\label{lastpage}
\end{document}